\documentclass[%
 reprint,
 amsmath,amssymb,
 prb,
]{revtex4-1}

\usepackage{graphicx}
\usepackage{dcolumn}
\usepackage{bm}
\usepackage{multirow}
\usepackage{color}

\newcommand{\specialcell}[2][c]{%
  \begin{tabular}[#1]{@{}c@{}}#2\end{tabular}}
  
\usepackage{float}

\begin{document}

\preprint{APS/123-QED}

\title{Electronic structure beyond the generalized gradient approximation for Ni$_2$MnGa}
\author{D.R.\ Baigutlin$^{1,2}$}
\author{V.V.\ Sokolovskiy$^{1}$}
\author{O.N.\ Miroshkina$^{1,2}$}
\author{M.A.\ Zagrebin$^{1,3}$}

\author{J.~Nokelainen$^{2}$}
\author{A.~Pulkkinen$^{2}$}
\author{B.\ Barbiellini$^{2,4}$}
\author{K.\ Pussi$^{2}$}
\author{E.\ L\"{a}hderanta$^{2}$}

\author{V.D.\ Buchelnikov$^{1,5}$}
\author{A.T.\ Zayak$^{6}$}
\affiliation{$^{1}$Faculty of Physics, Chelyabinsk State University, 454001 Chelyabinsk, Russia}
\affiliation{$^{2}$LUT University, FI-53851 Lappeenranta, Finland}
\affiliation{$^{3}$National Research South Ural State University, 454080 Chelyabinsk, Russia}
\affiliation{$^{4}$Department of Physics, Northeastern University, Boston, MA 02115, USA}
\affiliation{$^{5}$National University of Science and Technology "MISiS", 119049 Moscow, Russia}
\affiliation{$^{6}$Bowling Green State University, Bowling Green, OH 43403, USA}

\date{\today}

\begin{abstract}
The stability of the nonmodulated martensitic phase, the austenitic Fermi surface and the phonon dispersion relations for ferromagnetic Ni$_2$MnGa are studied using density functional theory.
Exchange-correlation effects are considered with various degrees of precision, starting from the simplest local spin density approximation~(LSDA), then adding corrections within the generalized gradient approximation~(GGA) and finally, including the meta-GGA corrections within the strongly constrained and appropriately normed~(SCAN). We discuss a simple procedure to reduce a possible overestimation of magnetization and underestimation of nesting vector in SCAN by parametrically decreasing self-interaction corrections.


\end{abstract}


\maketitle

\newpage

\section{Introduction}
During the last decades, the intermetallic Heusler alloys family has attracted enormous interest because a wide spectrum of the remarkable properties related to elasticity, magnetism, and thermodynamics~\cite{Vasilev2003, Buchelnikov2006, Entel2006, Entel2008, Planes2009, Entel2010, Entel2011, Hickel}.
In particular, ternary Heusler compounds with generic formula $X_2YZ$ crystallize in L2$_1$-cubic structure in the austenitic phase and can undergo a martenisitic phase transition~(MT) to lower symmetry structure upon cooling.  
In general, $X$ and $Y$ are 3$d$ transition metals and $Z$ is an $sp$ element of III-V group in the periodic table.
Usually, the martensitic temperature~($T_{\mathrm{M}}$) depends on the chemical composition, and thus on local interactions, plastic deformation, and heat treatment protocols~\cite{Planes2009, Buchelnikov2006, Vasilev2003}. 
For most compounds, $T_{\mathrm{M}} < T_{\mathrm{C}}$, where $T_{\mathrm{C}}$ is the Curie temperature.
As a result, the magnetic degrees of freedom are expected to strongly couple with the lattice, especially when $T_{\mathrm{M}}$ is close to $T_{\mathrm{C}}$~\cite{Bozhko1998, Vasilev1999, Vasilev2003, Khovaylo_2005}. 

In 1996, Ullakko and collaborators studied the magnetic control over the shape-memory effect in the ferromagnetic Ni$_2$MnGa Heusler alloy~\cite{Ullakko}.
These authors were able to induce reversible deformations of 0.2\% by application of magnetic fields below 1~T.
In 2002, a larger effect of about 10\% was obtained in similar compounds with a slightly different composition~\cite{Ullakko_2002}.
Today, a great effort is deployed to search new Heusler compounds with optimal properties for sensors and actuators~\cite{Kohl2007, Buchelnikov2006, Irzhak2014}.

Concerning the origin of the MT in Ni$_2$MnGa~\cite{Webster2}, a theoretical explanation based on the Jahn–Teller effect has been proposed by Fujii~\textit{et~al.}~\cite{Fujii_1989} from the calculated densities of states~(DOS), which is also consistent both with another DFT calculation by Ayuela~\textit{et~al.}~\cite{Ayuela_1999} and photoemission experiments by Opeil~\textit{et~al.}~\cite{Opeil}. 
However, the existence of three-layered premartensetic~(3M), martensitic five-layered~(5M), seven-layered~(7M) modulated phases in addition to the non-modulated tetragonal phase cannot be explained by the Jahn–Teller effect only. 
As suggested by several authors (see e.g. Refs.~\cite{Vasiliev-1990,Zheludev-1995,Zheludev-1996,Chernenko-1993}), the formation of modulated phases is associated with an anomalous transverse acoustic~TA$_2$ mode in the [110] direction at the wave vector $\mathbf{q} = \frac{2\pi}{a} [\xi\xi0]$ with $\xi \approx 0.33$ and, consequently, a softening of shear modulus $C'$.
Thus, the parent cubic phase becomes unstable  because the atomic planes can shuffle along the [110] direction and the premartensitic phase transition occurs at $T_P \approx 260$~K~\cite{Stuhr1997, Khovailo_2001}. In the case of MT, the transition temperature is $T_M \approx 320$~K while the softening of TA$_2$ phonon branch is more pronounced~\cite{Bungaro}, and it is shifted to $\xi \approx 0.43$.  
The role of the phonon anomaly in destabilizing the austenitic phase can be explained by the nesting of the austenitic Fermi surface~\cite{Haynes_2012} (FS) and the electron-phonon coupling~\cite{Bungaro,Wan-2005}.
The relationship between the premartensitic lattice softening with the generalized susceptibility singularities produced by the FS nesting has been reviewed  by Katsnelson~\textit{et~al.}~\cite{Katsnelson}.

For magnetic shape memory compounds, some important questions regarding this relationship remain  to be clarified despite a large amount of work (for instance, see Refs. \cite{Velikokhatnyi1999, Bungaro, Lee, Haynes_2012, Opeil, Siewert, Entel2008, Siewert_diss}) because of the lack of information on correlation effects present in martensitic ferromagnetic phases.
Nevertheless, despite correlation effects, the FS nesting could still play an important role in the instability of the austenitic phase upon cooling.
Velikokhatnyi and Naumov~\cite{Velikokhatnyi1999} have first explored the FS nesting by using density functional theory~(DFT) within local spin-density approximation~(LSDA)~\cite{Vosko, Perdew2} and concluded that the nesting vector parameter $\xi$ extracted from generalized susceptibility is approximately 0.42 justifying a MT to the 5M phase rather than the premartensitic transition to the 3M phase.
Lee~\textit{et al.}~\cite{Lee} with the same method observed a nesting vector in agreement with the experimental phonon anomaly~\cite{Zheludev-1995,Zheludev-1996} by renormalizing the value of the magnetic moment with the factor of 70\%, which corresponds to the magnetization at $T_{\mathrm{P}}$. 

Haynes~\textit{et al.}~\cite{Haynes_2012} conducted a comprehensive  study combining positron annihilation experiments  with linear muffin-tin orbital electronic structure calculations within LSDA~\cite{Barbiellini-LMTO}. 
These authors found that the peaks of the generalized susceptibility~$\chi(\mathbf{q})$ appear both in spin-up and spin-down channels.
Siewert~\textit{et al.}~\cite{Siewert, Siewert_diss} examined the effect of corrections beyond LSDA on $\chi(\mathbf{q})$ in the framework of the generalized gradient approximation~(GGA).
Their results confirm that the FS nesting is present not only for spin-down channel, but for spin-up as well.
They also noticed new susceptibility peaks along other directions produced by GGA.
Bungaro~\textit{et al.}~\cite{Bungaro} performed both phonon dispersion and FS calculations demonstrating that GGA corrections to LSDA are beneficial in order to improve the agreement between theory and experiment.  



Since correlation effects in Ni$_2$MnGa appear to play a crucial role on total energies, DOS, and magnetic moments~\cite{Buchelnikov-2019,Cococcioni-2012}, we also expect major corrections beyond LSDA and GGA for describing FS nesting and phonon instabilities.
A simple step forward is the strongly constrained and appropriately normed~(SCAN) functional, which is the most promising meta-GGA scheme due to the number of exact constrains fulfilled by the exchange-correlation energy~\cite{Perdew5, Tao, Sun}.
SCAN cures the unphysical interaction of an electron with itself, which occurs in LSDA and in GGA. The self-interaction correction (SIC) can be measured with a Coulomb energy $U$, which is related to the exchange correlation hole \cite{Barbiellini2005}. 
It is possible that for electrons at the Fermi surface, the SIC must be reduced \cite{Norman} because of the itinerant character of the wave functions. 
The correction must in fact vanish in the limit of plane-waves.
To examine this possible SCAN over-correction, we also consider a SCAN$-U$ scheme, where the SIC is reduced by an amount $U$ for the $3d$ orbitals on Mn atoms. 
This parametric study SCAN$-U$ can provide a first understanding how to further improve SCAN for predicting functional Heusler alloys and their magnetic properties accurately.

The outline of the paper is as follows. 
Section~II contains the computational methods and calculation details. 
Section~III is devoted to the discussion of the results of exchange correlation corrections on FS, nesting vector, generalized susceptibility, and phonon dispersion relations.
The concluding remarks are presented in Section~IV.
    
\section{Calculation details}
The electronic structure calculations were performed using the spin-polarized DFT within the projected augmented wave~(PAW) method implemented in Vienna Ab-initio Simulation Package~(VASP)~\cite{VASP1, VASP2}.
LSDA and GGA parametrized by Perdew-Zunger~\cite{PZ-LDA} and  Perdew-Burke-Ernzerhof~\cite{Perdew4}, respectively, were used to describe correlation energy. 
Effects beyond GGA are described  with the meta-GGA corresponding to the SCAN implementation~\cite{Sun}.
A parametric study was performed using SCAN$-U$ method.
The $k$-points within the Brillouin zone were generated using a uniform Monkhorst-Pack~\cite{Monkhorst-Pack} mesh of 12$\times$12$\times$12. 
The cutoff energy was set to 700~eV.
The PAW pseudopotentials were generated with the following atomic configurations: Mn($3p^63d^64s^1$), Ni($3p^63d^94s^1$), and Ga($3d^{10}4s^24p^1$).
The calculations were converged with the energy accuracy of $10^{-6}$~eV/atom.
Conjugate gradient algorithm was used to minimize all the residual forces until the convergence criteria of 0.01~eV/\AA.


For the FS modeling, the $51\times51\times51$ Monkhorst-Pack $k$-grid was used.
The generalized susceptibility was calculated for both majority and minority spin channels as following~\cite{Velikokhatnyi1999, Bungaro, Lee}

%
\begin{equation}\label{int_begin} 
    \chi(\mathbf{q}) = \sum\limits_{n,m,\mathbf{k}} \frac{f(\varepsilon_{m}(\mathbf{k}) [1-f(\varepsilon_{n}(\mathbf{k}))]}{\varepsilon_{n}(\mathbf{k}+\mathbf{q}) - \varepsilon_{m}(\mathbf{k})},
\end{equation}
where $f(\varepsilon_{m(n)})$ is the Fermi-Dirac distribution function and $\varepsilon_{m}(\textbf{k})$, $\varepsilon_{n}(\textbf{k})$ are energies corresponding to the $m$ and $n$ band at the wave vector \textbf{k}. 
The peaks of $\chi(\mathbf{q})$ indicate electronic instabilities associated with the FS nesting.

Phonon calculations along a single direction  were performed using the PHONON~\cite{Parlinski-software,Parlinski-1997} package.
We used a 48-atom supercell consisting of three 16-atom initial tetragonal cells  merged in the direction [110] of the tetragonal cell.
The PHONON package uses Hellmann-Feynman forces obtained  within GGA and SCAN.
The force constants were calculated by displacing each atom along the cartesian coordinates ($x$, $y$, and $z$) by $\pm$0.03~\AA. 

\section{Results and discussions}

The geometry optimization and ground state properties calculations were performed for  cubic L2$_1$ ($Fm \bar{3}m$, No.~225) and tetragonal  L1$_0$ ($Fmmm$, No.~69) structures of Ni$_2$MnGa.
Calculations were performed on four atoms cells both for austenitic and martensitic phases.
For L2$_1$ structure, Ni atoms occupy 8$c$~((1/4, 1/4, 1/4) and 3/4, 3/4, 3/4)) Wyckoff positions, while Mn and Ga atoms site at 4$b$~(1/2, 1/2, 1/2) and 4$a$~(0, 0, 0), correspondingly.
For L1$_0$, Ni atoms site on 8$f$~((1/4, 1/4, 1/4) and 3/4, 3/4, 3/4)) Wyckoff positions, Mn atom locates at 4$b$~(0, 0, 1/2), and Ga atom sites at 4$a$~(0, 0, 0).
The results of geometry optimization for martensitic and austenitic structures obtained with LDA, GGA, SCAN, and SCAN$-U$ are presented in Table~\ref{table1}.

\begin{table}[h!tb]
    \caption{The equilibrium lattice parameters~$a_0$ and $a_{t}$ in \AA, total magnetic moment~$\mu_{tot}$ in $\mu_B$/f.u. for austenitic and martensitic structure and energy difference between austenite and martensite~$\Delta E$ in meV/atom of Ni$_2$MnGa calculated with LSDA, GGA, SCAN, and SCAN$-U$~($U$= 1 and 1.8~eV). For comparison, available experimental data are presented.}
    \def\arraystretch{1.1}
 \resizebox{\columnwidth}{!}{%
    \begin{tabular}{lcccccccccccc}
    \hline \hline
    & \multicolumn{4}{c}{Austenite}               & \multicolumn{6}{c}{Martensite}      & \multirow{2}{*}{$\Delta E$} \\ \cline{2-4} \cline{6-10} 
    & $a_0$  & & $\mu_{tot}$ &      & $a_t$  & & $c/a$  & & $\mu_{tot}$ & &       \\
    \hline
    LDA & 5.633 & & 3.717 & & 5.207 & & 1.264  & & 3.821  & & 16.8      \\
    GGA & 5.807 & & 4.137 & & 5.384  & & 1.250  & & 4.105  & & 8.1       \\ 
    SCAN    & 5.726 & & 4.726 & & 5.378  & & 1.213  & & 4.585  & & 11.9      \\ 
    \specialcell{SCAN$-U$ \\ (1.0 eV)}    & 5.706 & & 4.464 &  & 5.368  & & 1.198  & & 4.404  & & 19.8 \\ 
    \specialcell{SCAN$-U$ \\ (1.8 eV)}    & 5.690 & & 4.173 &  & 5.308  & & 1.230  & & 4.081  & & 28.1 \\ 
    Exp.              & 5.825\footnote{Ref.~\cite{Webster2} } & & \specialcell{3.63\footnote{Ref.~\cite{Ooiwa1992} (at 230 K)}} & & 5.52\footnote{Ref.~\cite{Martynov}} & & \specialcell{1.18$\pm$0.02\footnote{Ref.~\cite{Chernenko-1998}}, \\ 1.2\footnote{Ref.~\cite{Martynov}}} & & \specialcell{4.23\footnote{ Ref.~\cite{Ooiwa1992} (at 4.2 K)}} & & -  \\ \hline \hline

    \end{tabular}
    }
    \label{table1}
\end{table}


\subsection{Fermi surface and generalized susceptibility}

\begin{figure}[!t]
\includegraphics[scale=0.4]{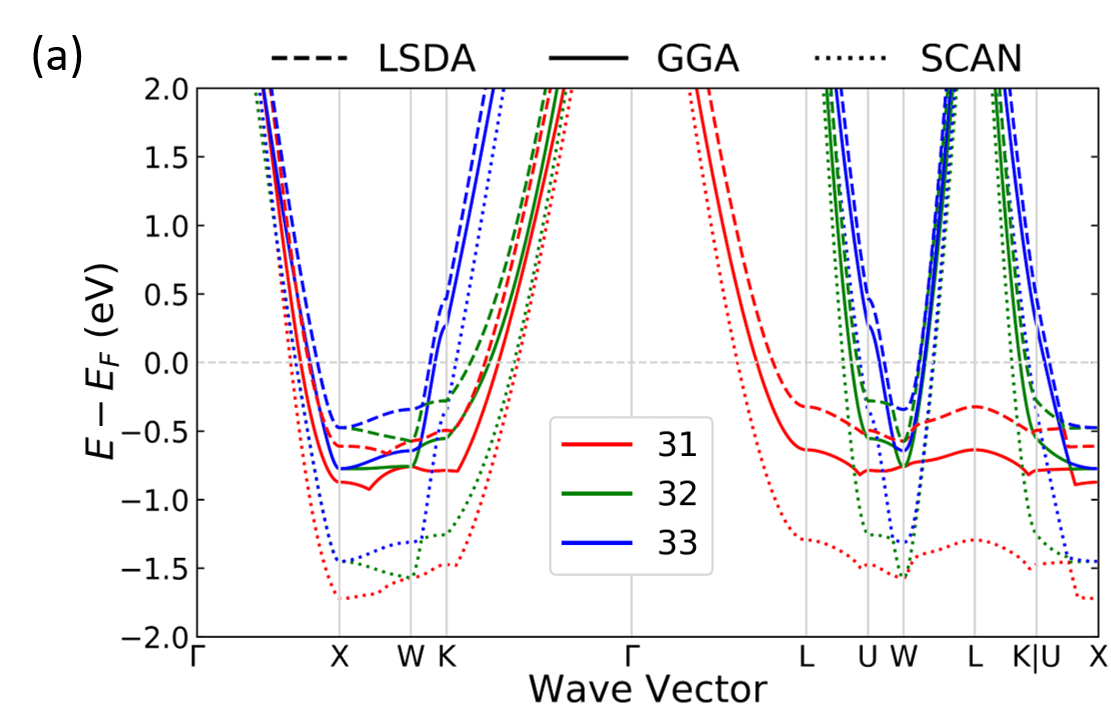}
\includegraphics[scale=0.4]{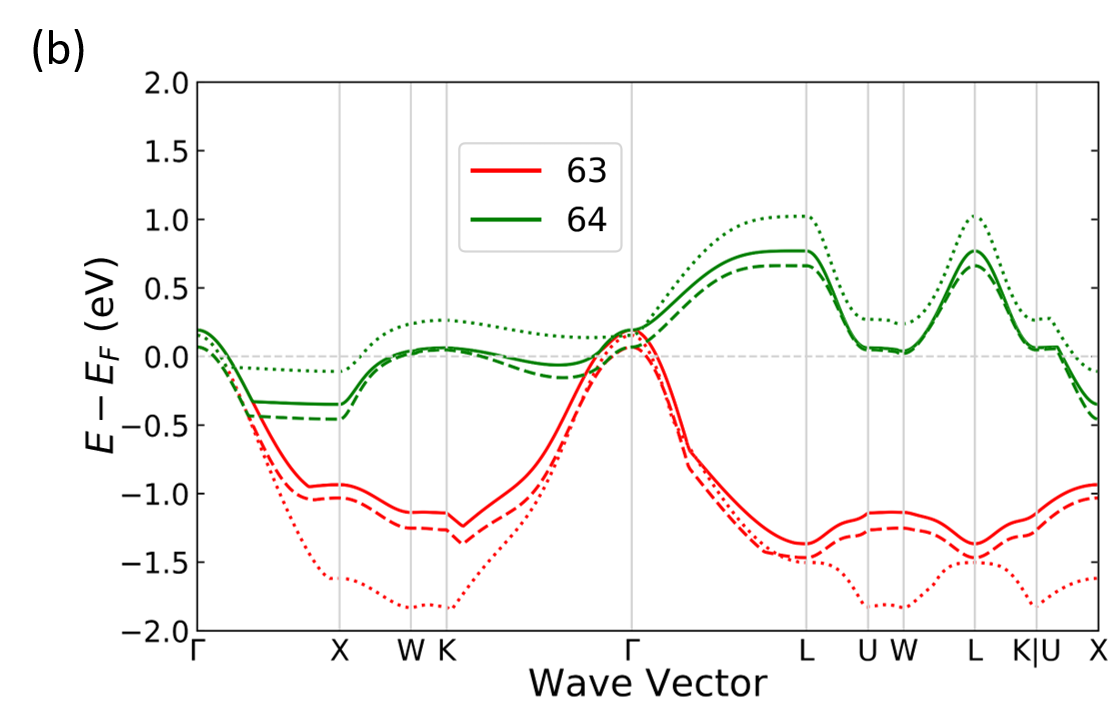}
\caption{ (a)~Spin-up and (b)~spin-down bands crossing the Fermi level. The dash, full, and dot lines correspond to the LSDA, GGA, and SCAN, respectively. }
\label{fig:fig1}
\end{figure}

Figure~\ref{fig:fig1} shows the band structure in the vicinity of Fermi level calculated at equilibrium volume of the austenitic phase.
The Fermi level is crossed by three spin-up (numbered as 31, 32, and 33) and two spin-down (numbered as 63 and 64) bands for LSDA, GGA, and SCAN.
The FS crossings appear to be robust toward the description of correlation but some differences can be observed.
The most significant difference is found along L-U-W and L-K-X paths for the spin-down bands as shown in Fig.~\ref{fig:fig1}(b). 
In this case, band 64 calculated within SCAN does not cross the Fermi level along the path K-$\Gamma$ in contrast to LSDA and GGA indicating that some parts of the FS vanish.

\begin{figure*}[!ht]
{
\includegraphics[scale=0.35]{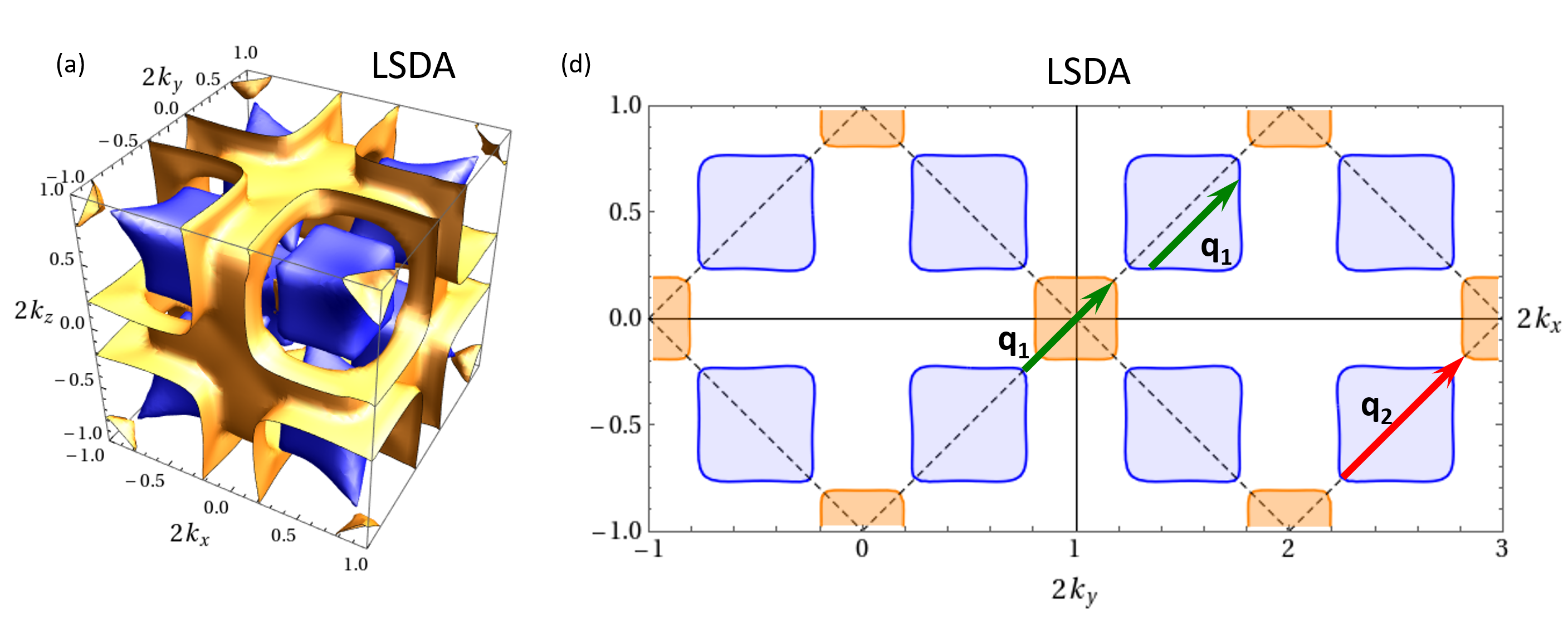}
\includegraphics[scale=0.35]{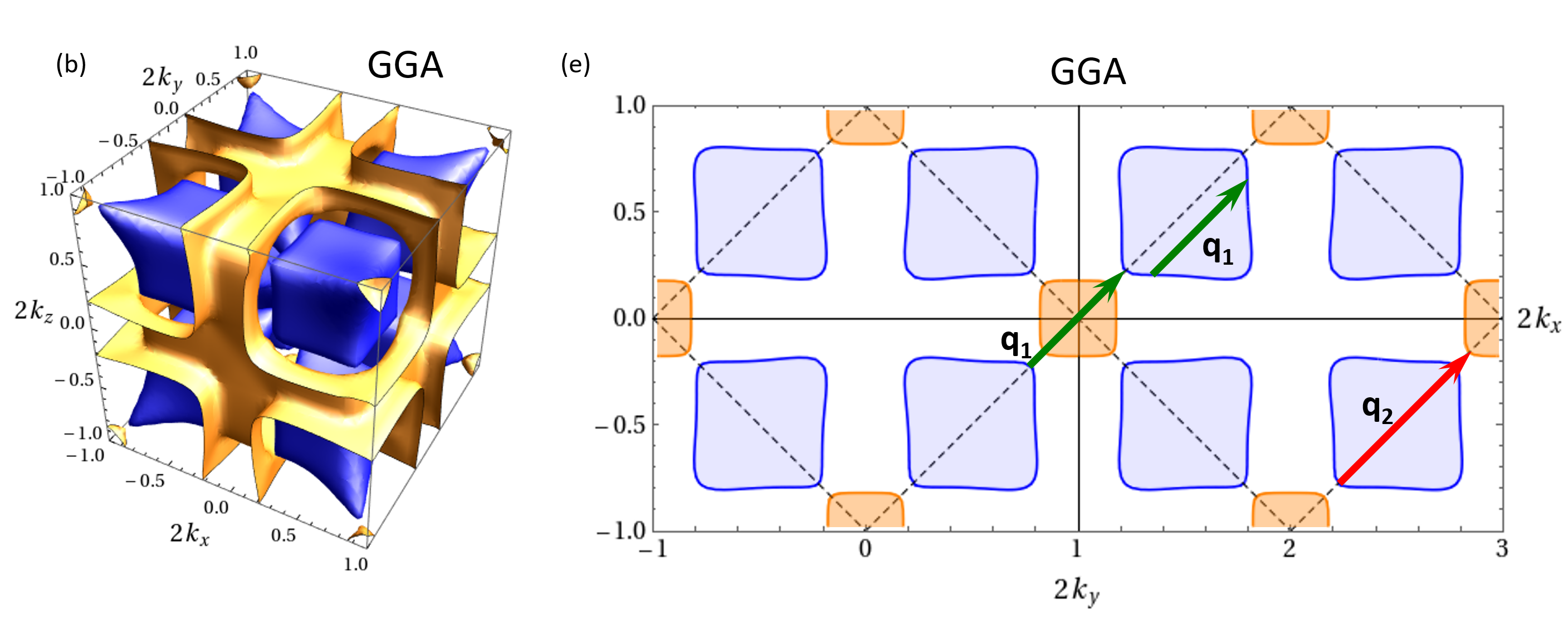}
\includegraphics[scale=0.35]{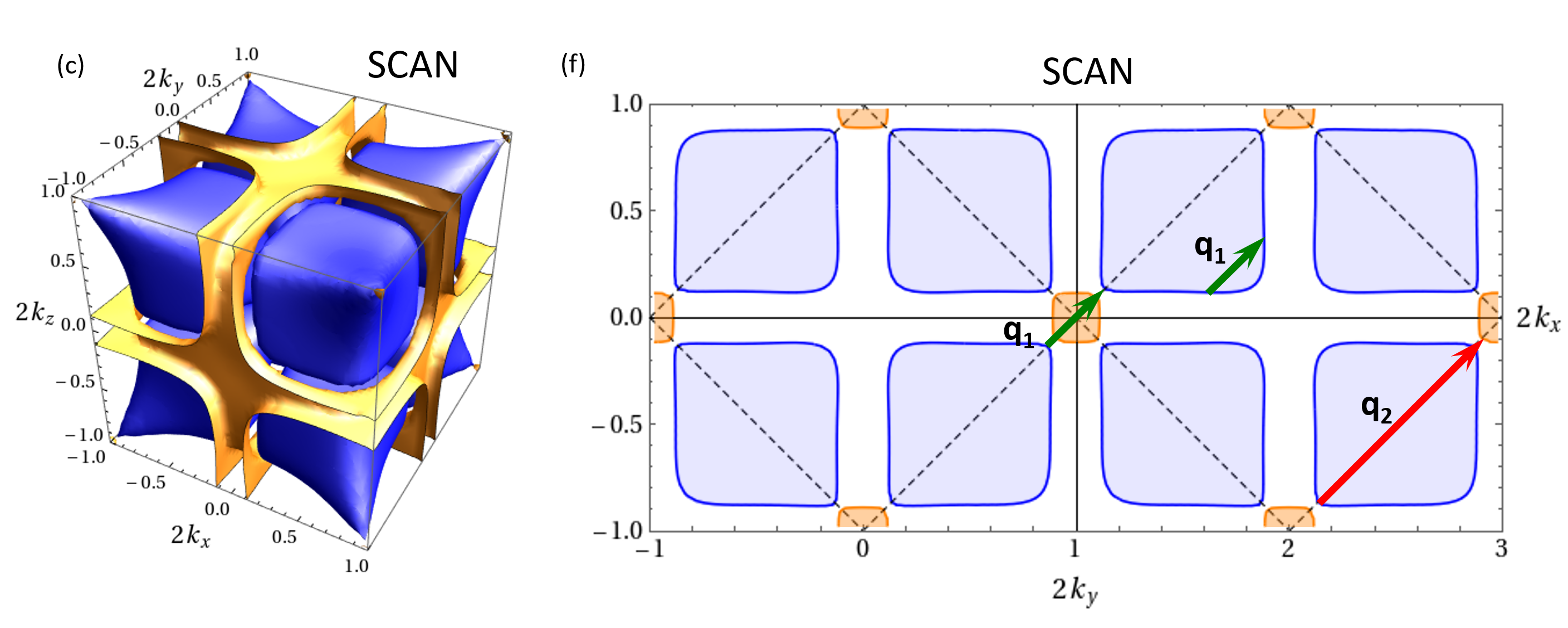}
}
\caption{ (Left panel) Fermi surfaces of minority spin band of Ni$_2$MnGa. Blue (orange) color corresponds to 63 (64) band in Fig.\ref{fig:fig1}, respectively. (right panel) Cross sections of the FS at $k_z = 0.5$($2\pi/a$). Red arrows illustrate the nesting vectors. Dot lines delineate the boundaries of the first Brillouin zone. }
\label{fig:fig2}
\end{figure*}

Figures~\ref{fig:fig2}~(a-c) show the minority spin FS for the austenitic phase within LSDA, GGA, and SCAN, which consists of two sheets corresponding to bands 63 and 64.
The results for LSDA and GGA are in good agreement with previous studies\cite{Siewert_diss, Haynes_2012}.
As a consequence of the increased magnetic moment in SCAN, a remarkable modification of the minority FS occurs.
In particular, the FS sheet associated to band 63 expands, while the sheet corresponding to band 64 shrinks.
Moreover, the disappearance of a FS piece located at the corner of the Brillouin zone is consistent with Fig.~\ref{fig:fig1}(b).

Figures~\ref{fig:fig2}~(d-f) present a visual analysis of the FS cross section at $k_z = 0.5$($2\pi/a$) in terms of nesting vectors $\textbf{q}_1$ and $\textbf{q}_2$.
These vectors correspond to peaks of  $\chi(\textbf{q})$ along the [110] direction. 
Our results give $|\textbf{q}_1|\approx 0.43$ and  0.39 for LSDA and GGA, respectively, which are in agreement with Lee~\textit{et al.}~\cite{Lee} and Bungaro~\textit{et al.}~\cite{Bungaro}.
The SCAN value is $|\textbf{q}_1| \approx 0.20$ reflecting significant differences in the FS nesting properties.

\begin{figure}[htb!]
    \begin{center}
        \includegraphics[scale=0.5]{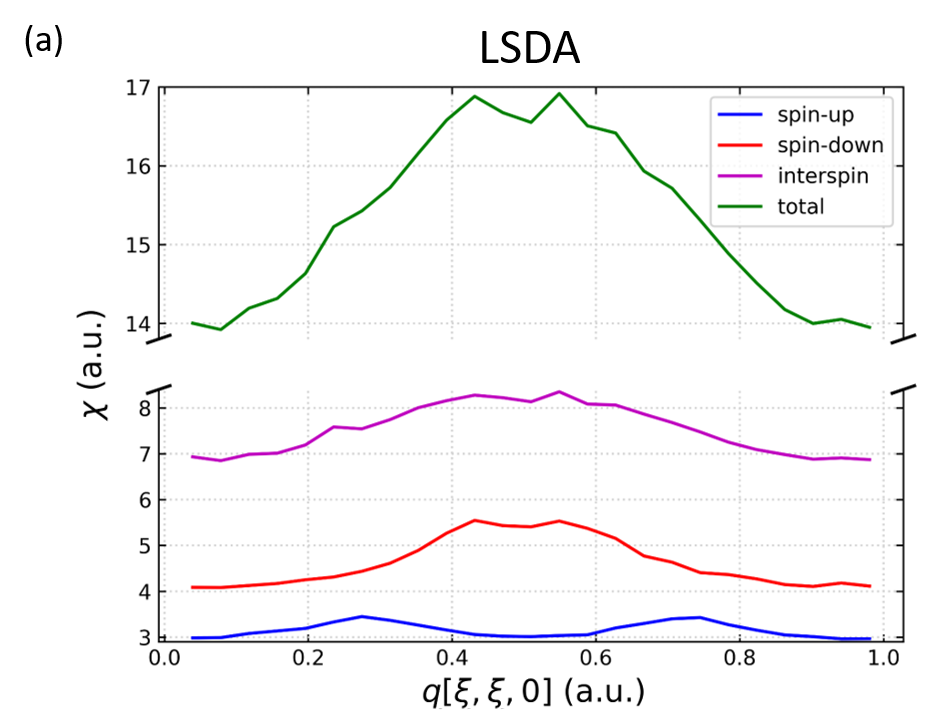}
        \includegraphics[scale=0.5]{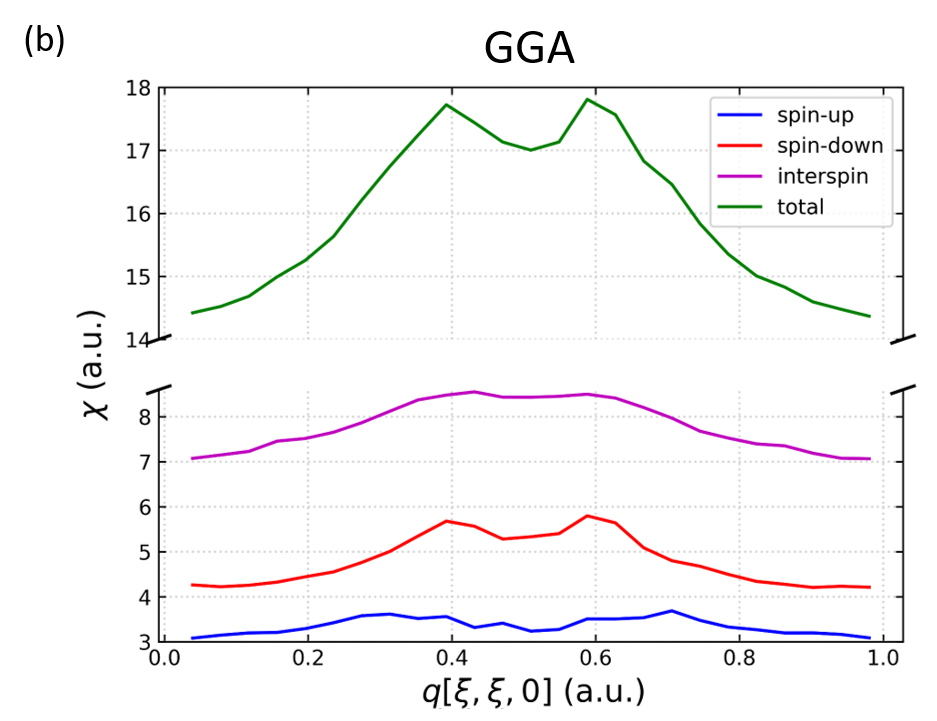}
        \includegraphics[scale=0.5]{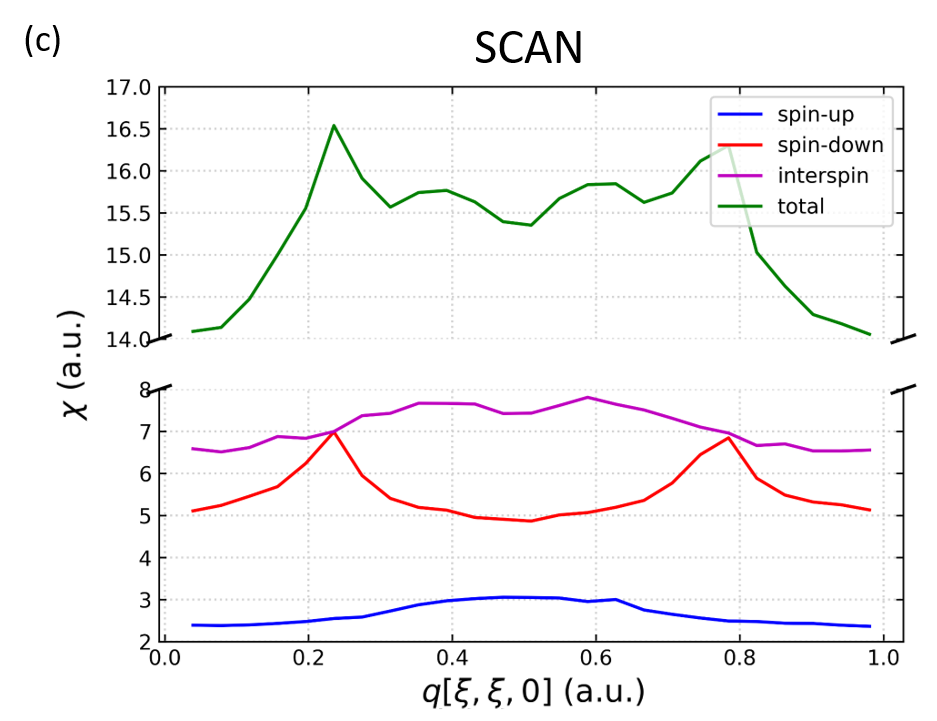}
    \end{center}
    \caption{ Cross sections of the generalized electron susceptibility along the [110] direction. The results are presented for spin-up, spin-down, interspin, and total contributions. }
    \label{fig:fig3}
\end{figure}

A more quantitative analysis of FS nesting properties visualizes the $\chi(\textbf{q})$ cross-section in the direction of [110] as shown in Fig.~\ref{fig:fig3}.
These spectra are presented for 
spin-up (contribution from the majority spin only), spin-down (contribution from the minority spin only), interspin (contribution from the interactions between spin-up and spin-down bands only), and total (contributions from spin-up, spin-down, and interspin) contributions.
Full 2D-maps of generalized electronic susceptibility  for the (110) plane are also illustrated in SM~\cite{SM}. 
According to the Fig.~\ref{fig:fig3}, the results look similar for the minority and majority contributions calculated with LSDA and GGA.
It appears that LSDA slightly reduces the distance between two susceptibility peaks with respect to GGA. 
The total contribution does not differ significantly for LSDA and GGA. 
The actual values of $|\textbf{q}_1|$ are 0.435 and 0.394 in LSDA and GGA, respectively, while for $|\textbf{q}_2|$ the values are 0.550 in LSDA and 0.596 in GGA. The GGA results are in good agreement with the previous studies~\cite{Velikokhatnyi1999, Lee, Siewert_diss}.
SCAN alters considerably the form of the generalized electronic susceptibility in Fig.~\ref{fig:fig3}.
This modification results in a significant separation of the two susceptibility peaks and the appearance of the additional peaks at $\textbf{q}_3$ and $\textbf{q}_4$ originating from interspin contribution.
Moreover, according to the profile shown in Fig.~\ref{fig:fig3}, very different nesting vectors $|\textbf{q}_1|$ and $|\textbf{q}_2|$ are found.
The values of their norms are $|\textbf{q}_1| = 0.263$ and $|\textbf{q}_2| = 0.784$.
These results show that the correlation effects bring new singularities in $\chi(\textbf{q})$, which can explain a more complex landscape of the competing nonmodulated and modulated phases as in the case of stripes in cuprate superconductors~\cite{zhang2018}. 
Thus, SCAN modifies both spin-up and spin-down $\chi(\textbf{q})$ contributions.

\subsection{Reduction of SIC}
\begin{figure}[!htb]
    \includegraphics[scale=0.55]{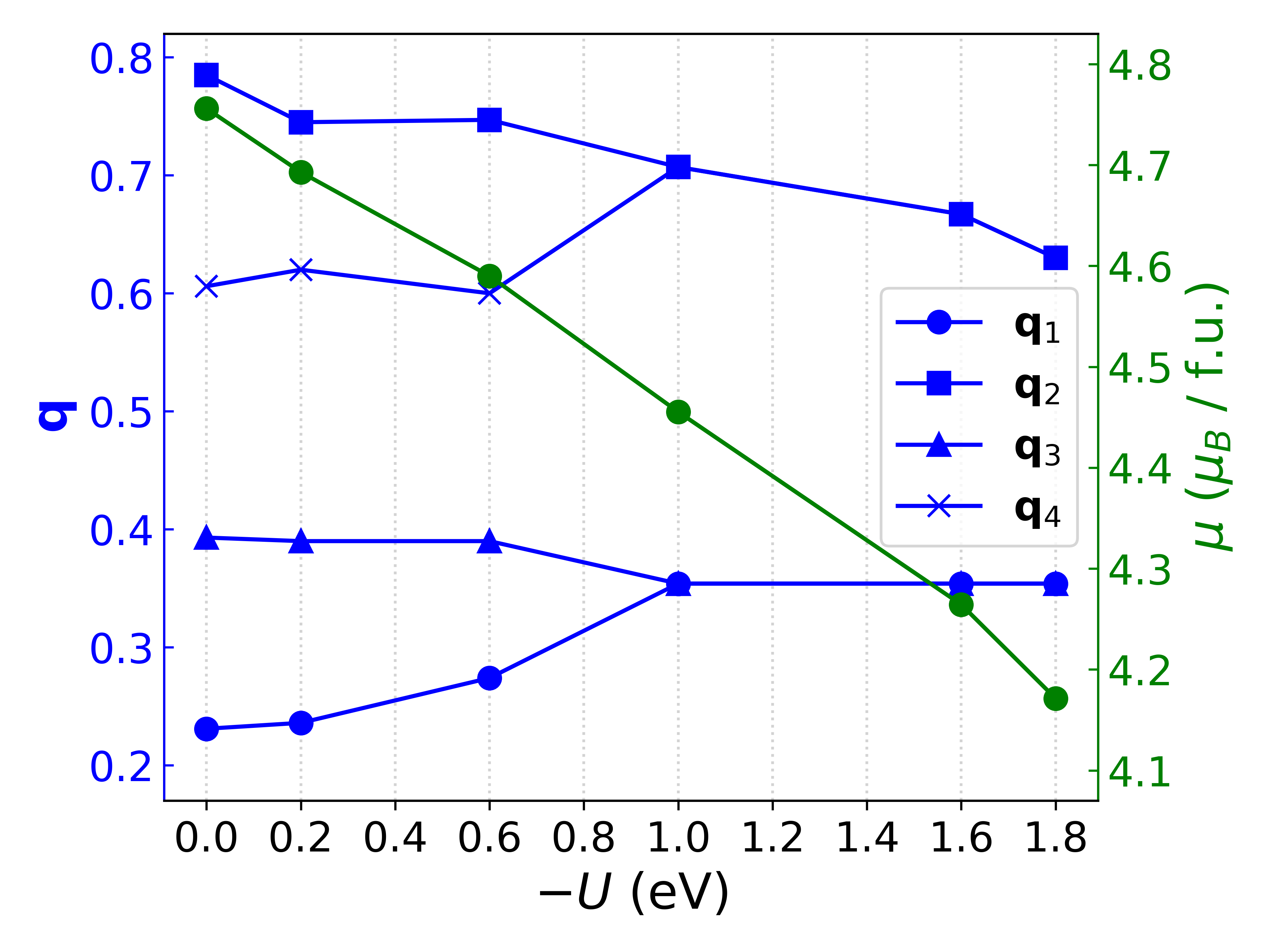}
    \caption{ Nesting vector (left axis) and total magnetic moment (right axis) as a function of the parameter~$-U$ controlling the SIC .} 
    \label{fig:fig4}
\end{figure}

According to several authors~\cite{Trickey, Barbiellini2005}, SCAN exaggerates the magnetic moment in some 3$d$ transition metals~\cite{Isaacs-2018,Ekholm-2018}.
The deorbitalization of the SCAN potential has been recently proposed to address this problem~\cite{Trickey}.
Another simple way to reduce SIC is to use a SCAN-$U$ method where $U$ is a parameter measuring the excess of SIC.
Figure~\ref{fig:fig4} shows a parametric study of the SCAN SIC reduction. 
When $U=0$, we have full SIC, while a finite $U$ measures the suppression of SIC.

As expected, the FS shape becomes more similar to the GGA case when a larger value of $U$ is removed from SCAN (see SM~\cite{SM}).
At the same time, the magnetic moment decreases and approach the GGA value of 4.17~$\mu_B$ when $U \approx 2$~eV.
Concerning $\chi(\textbf{q})$, the reduction of SIC leads to the merging of the peaks $(\textbf{q}_1, \textbf{q}_3)$ and $(\textbf{q}_2, \textbf{q}_4)$ when the critical value $U = 1$~eV is reached.
In this case, the 2D-map shown in SM~\cite{SM} becomes similar to the GGA 2D-map.
From this parametric study, we deduce that SIC associated with SCAN corresponds to an effective GGA$+U$ scheme with $U \approx 2$~eV.
This value of $U$ is significantly lower than the one proposed by Himmetoglu~\textit{et~al.}~\cite{Cococcioni-2012} ($U=5.97$~eV) and  the one proposed by \c{S}a\c{s}{\i}o\u{g}lu~\textit{et al.}~\cite{Sasiouglu} ($U \approx4$~eV for Mn atom).
 One should keep in mind that within the GGA+$U$ method, increasing $U$ on Mn sites results in a stronger SIC for the $3d$-Mn orbitals. 
Several authors~\cite{koubsky2018, bodnarova2020, zeleny2020} have shown that the parameter $U$ is needed to obtain better agreement with experiments for elastic constants, tetragonal ratios, and magnetic anisotropy energy of Ni$_2$MnGa. In particular, Refs.~\cite{bodnarova2020, zeleny2020} have shown that $U$ should be 1.8~eV for Mn. This value is smaller than the effective Coulomb correlation of 2~eV in SCAN, therefore in order to improve the agreement with experiment, one needs to consider a revised SCAN scheme where the SIC is reduced at the Mn sites.  
In our case, the use of $–U$ mimics this SIC reduction. 
SIC reductions for systems with partially filled $3d$ bands has been discussed by Barbiellini and Bansil~\cite{Barbiellini2005}.

\subsection{Phonons}

In the case of the austenitic phase, Bungaro~\textit{et al.}~\cite{Bungaro} showed that GGA gives imaginary phonon frequencies at zero temperature.
In SCAN, the appearance of the additional  peaks in $\chi(\textbf{q})$ makes this phase even more unstable. This can lead to softening of additional acoustical modes.

\begin{figure}[t]
    \includegraphics[scale=0.6]{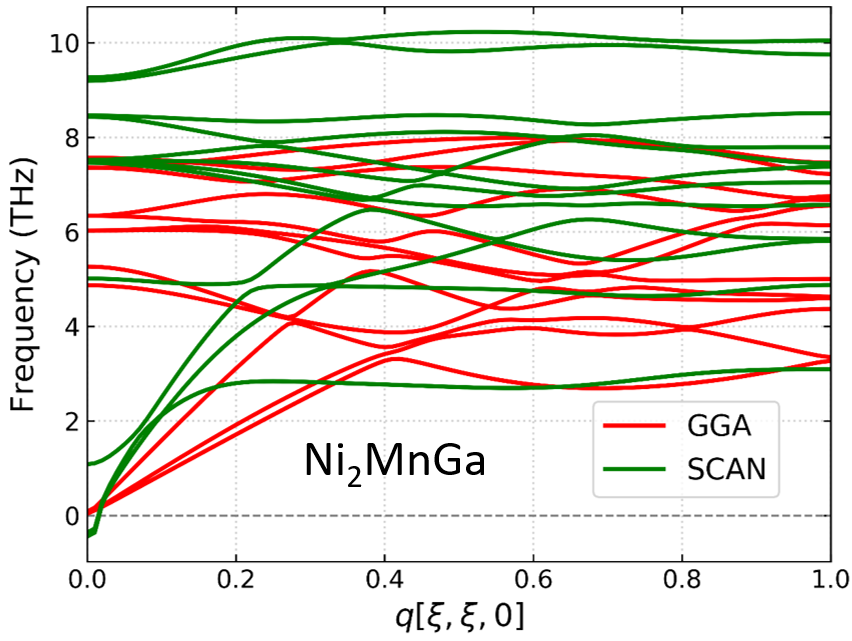}
    \caption{Phonon dispersion in the nonmodulated martensitic phase for GGA~(red) and SCAN~(green). The SCAN dispersion presents an imaginary excursion near the $\Gamma$ point. }
    \label{fig:fig5}
\end{figure}

Figure~\ref{fig:fig5} shows the phonon dispersion in the case of nonmodulated martensitic L1$_0$ phase.
The GGA phonon dispersion yields real frequencies for all phonon wave vectors in agreement with the calculation by Zayak~\textit{et al.}~\cite{Zayak2003}.
The SCAN calculation reveals an unstable mode near the $\Gamma$ point as shown in Fig.~\ref{fig:fig5}. Such small instability could be due by the fact that the landscape of almost degenerate solutions within SCAN usually becomes very complex as shown by Zhang~\textit{et al.}~\cite{zhang2018} therefore SCAN calculations are much more difficult to relax to the groundstate structure compared to the GGA.  Interestingly, Himmetoglu~\textit{et al.}~\cite{Cococcioni-2012} claim that the ground state is a modulated martensite. Moreover, even within GGA, Zelen\'y \textit{et al.}~\cite{Zeleny-2016} indicated that the 4O martensitic structure is 2~meV/atom below nonmodulated martensite solution.
%


\section{Conclusion}
The concept of heterogeneity is important to explain large magnetic shape memory effects~\cite{Aaron_diss} in Ni$_2$MnGa, where new phases inside samples are themselves a heterogeneous modulation of their parent phase~\cite{lookman2009}. These modulated phases~\cite{Niemann2012} can be stabilized within advanced DFT approaches beyond GGA. Competing inhomogeneous orders are a central feature of correlated electron materials, including the high-temperature superconductors. For example, by using schemes beyond GGA in YBa$_2$Cu$_3$O$_7$~\cite{zhang2018}, new landscape of solutions characterized by stripe orders with large magnetic moments on Cu atoms have been recently uncovered. Similarly, one has found that the energy minimization with the same DFT scheme beyond the GGA is controlled by large Mn local moments in elemental manganese~\cite{pulkkinen2020}. These observations demonstrate that correlation effects enable a new generation of understanding of the large magnetic shape memory effect, and how this property emerges through the interplay of spin, charge, and lattice degrees of freedom.

Our results indicate that SCAN behaves as an effective GGA$+U$ scheme with $U$ parameter of about 2~eV.
This amount of SIC captured by SCAN modifies the nesting properties of the austenitic FS and thereby creates new singularities in the electron susceptibility~$\chi(\textbf{q})$ justifying a more complex phase diagram.
Consequently, the nonmodulated martensitic phase becomes unstable within SCAN as demonstrated by the softening of some calculated phonon modes. 
Moreover, the total magnetic moment increases significantly as in other $3d$ transition metals such iron~\cite{Ekholm-2018}.
This magnetic enhancement has been considered exaggerated in the literature.
Therefore, it is possible that SIC corresponding to $U = 1.8$~eV  should be renormalized toward the critical value $U=1$~eV, where $\chi(\textbf{q})$ starts to develop new singularities due to correlation effects. 
From the present results, we conclude that the amount of SIC in Ni$_2$MnGa can be measured by an effective $U$ in the interval of $1-2$~eV.

\vspace{0.5cm}
\section{Acknowledgement}
This work was supported by Russian Science Foundation No.~17-72-20022. B.B.~acknowledges support from the COST Action CA16218. V.B. acknowledges support from the NUST "MISiS" No.~K2-2020-018

\bibliography{main}

\end{document}



\preprint{APS/123-QED}

\title{Electronic structure beyond the generalized gradient approximation for Ni$_2$MnGa}
\author{D.R.\ Baigutlin$^{1,2}$}
\author{V.V.\ Sokolovskiy$^{1}$}
\author{O.N.\ Miroshkina$^{1,2}$}
\author{M.A.\ Zagrebin$^{1,3}$}

\author{J.~Nokelainen$^{2}$}
\author{A.~Pulkkinen$^{2}$}
\author{B.\ Barbiellini$^{2,4}$}
\author{K.\ Pussi$^{2}$}
\author{E.\ L\"{a}hderanta$^{2}$}

\author{V.D.\ Buchelnikov$^{1,5}$}
\author{A.T.\ Zayak$^{6}$}
\affiliation{$^{1}$Faculty of Physics, Chelyabinsk State University, 454001 Chelyabinsk, Russia}
\affiliation{$^{2}$LUT University, FI-53851 Lappeenranta, Finland}
\affiliation{$^{3}$National Research South Ural State University, 454080 Chelyabinsk, Russia}
\affiliation{$^{4}$Department of Physics, Northeastern University, Boston, MA 02115, USA}
\affiliation{$^{5}$National University of Science and Technology "MISiS", 119049 Moscow, Russia}
\affiliation{$^{6}$Bowling Green State University, Bowling Green, OH 43403, USA}

\maketitle


\section{Fermi surface}

\begin{figure}[!h]
    \includegraphics[scale=0.33]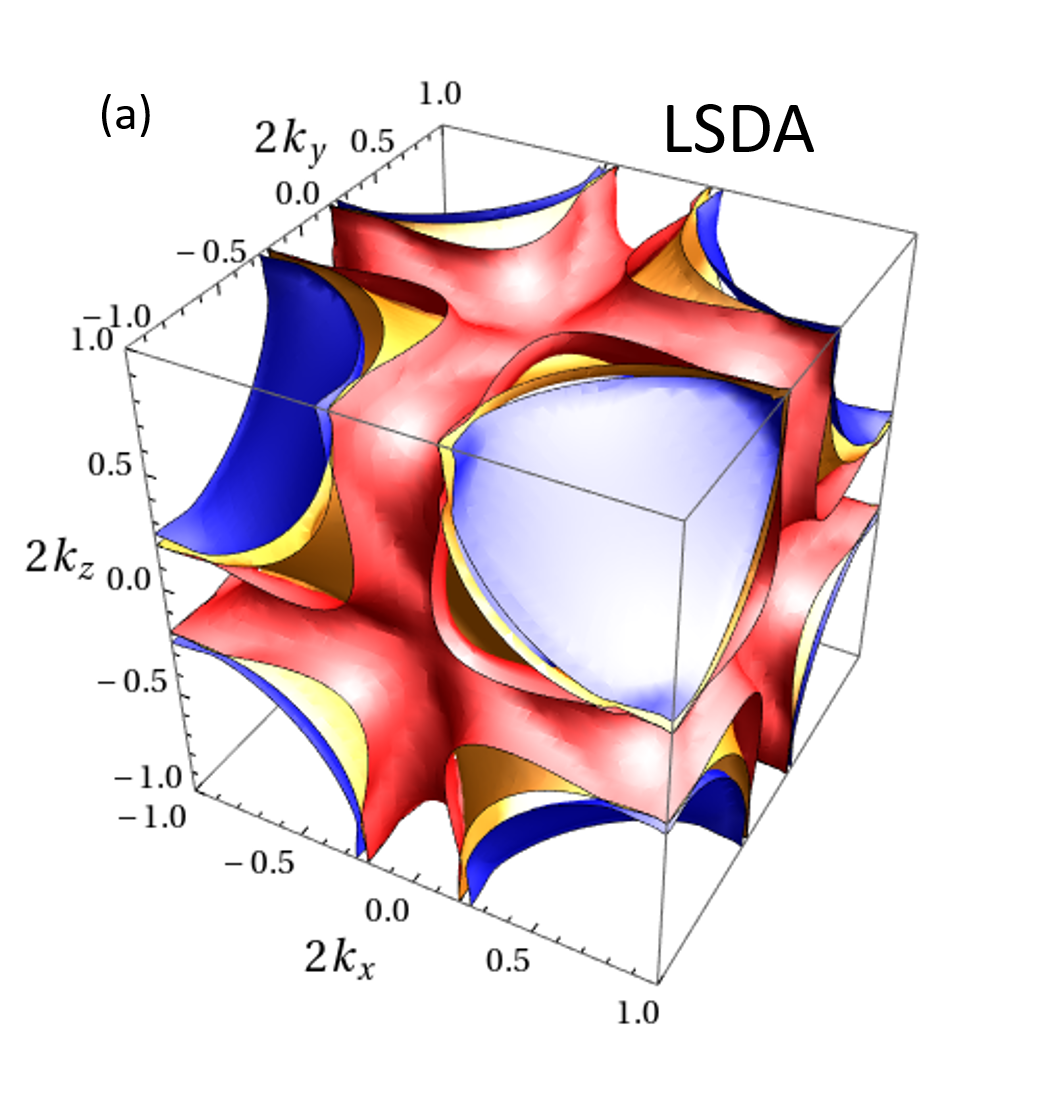}
    \includegraphics[scale=0.33]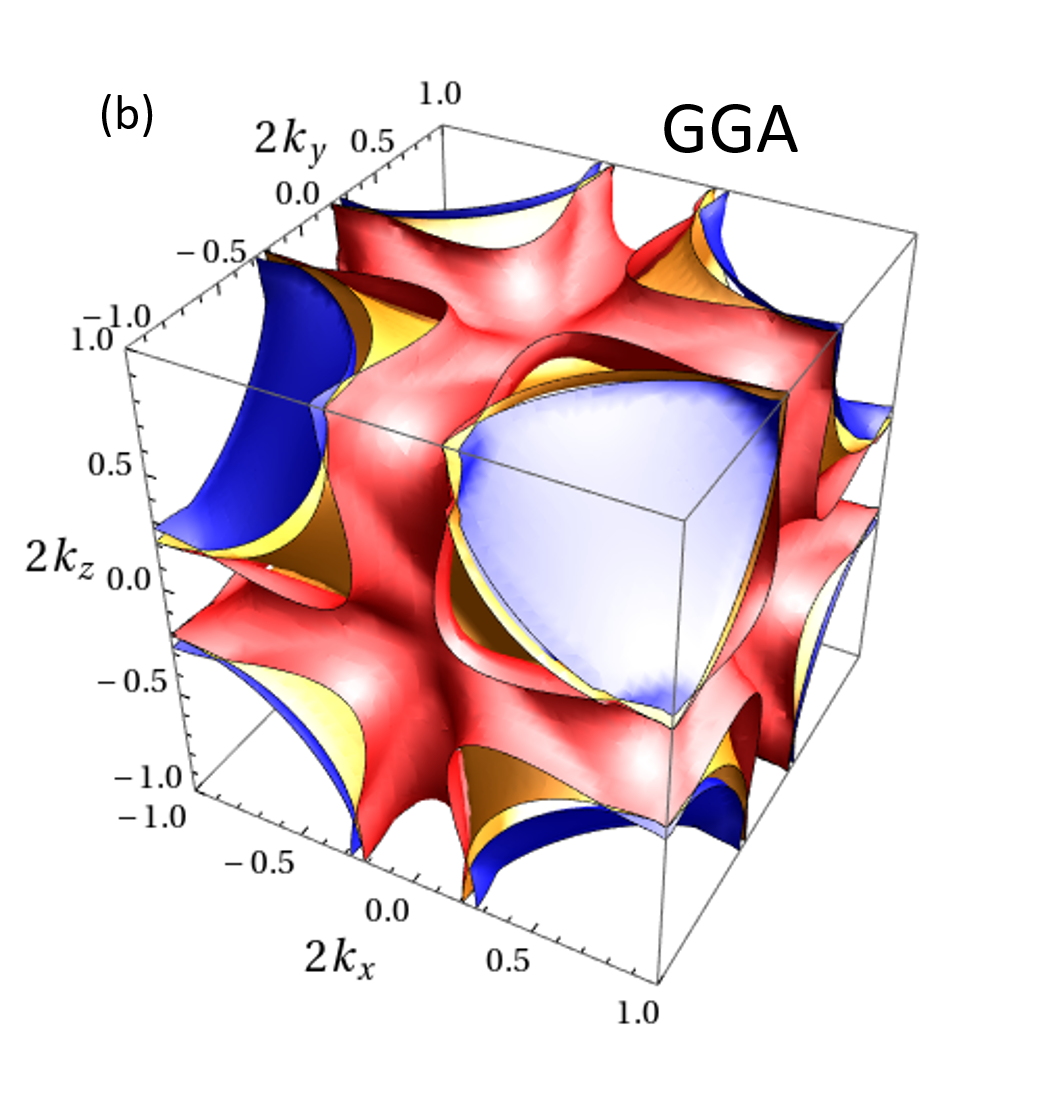}
    \includegraphics[scale=0.33]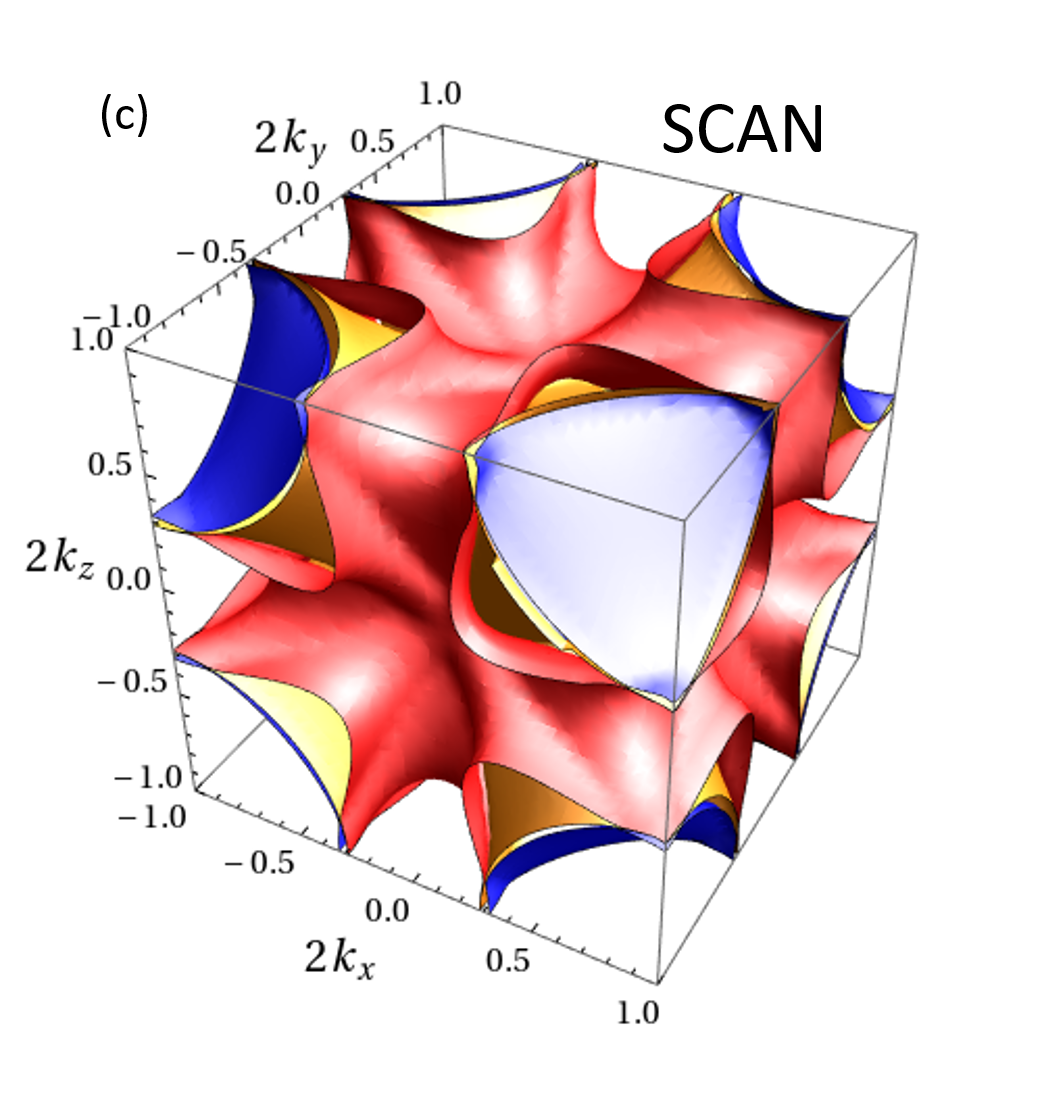}

    \caption{ Fermi surface for spin-up electrons for an austenitic structure of Ni$_2$MnGa. }
    \label{fig:fig1_s}
\end{figure}

Fig.~S\ref{fig:fig1_s} shows the Fermi surface for spin-up electrons calculated in the LSDA, GGA, and SCAN functionals.
It is seen, the Fermi surface of the spin-up bands does not change significantly, band 31 (red) expands mainly.

\section{Generalized electronic susceptibility}

\begin{figure*}[!t]
    \includegraphics[scale=0.33]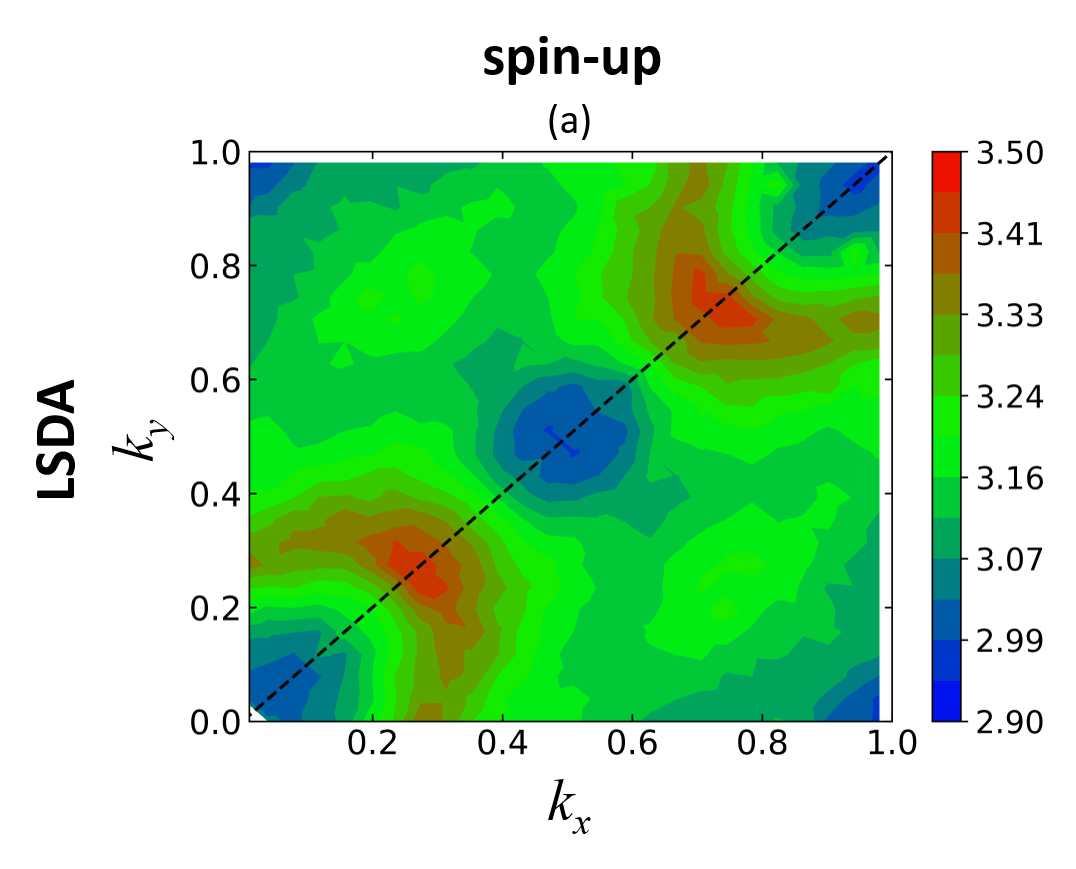}
    \includegraphics[scale=0.33]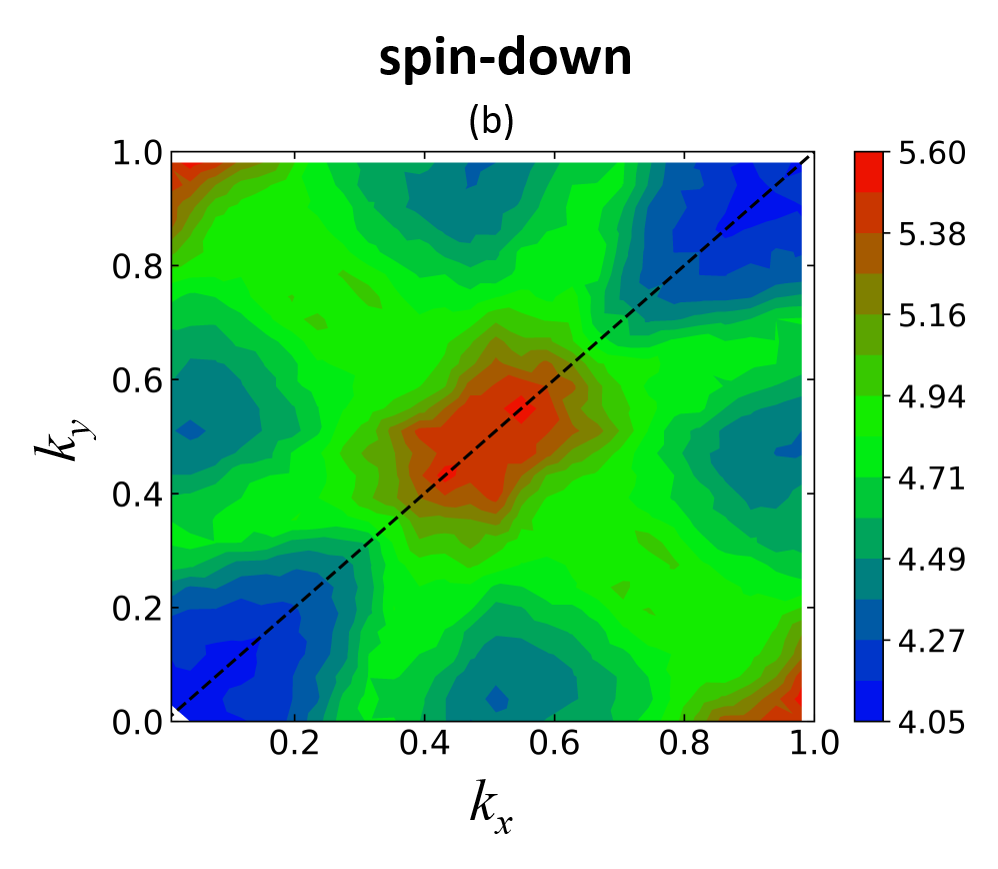}
    \includegraphics[scale=0.33]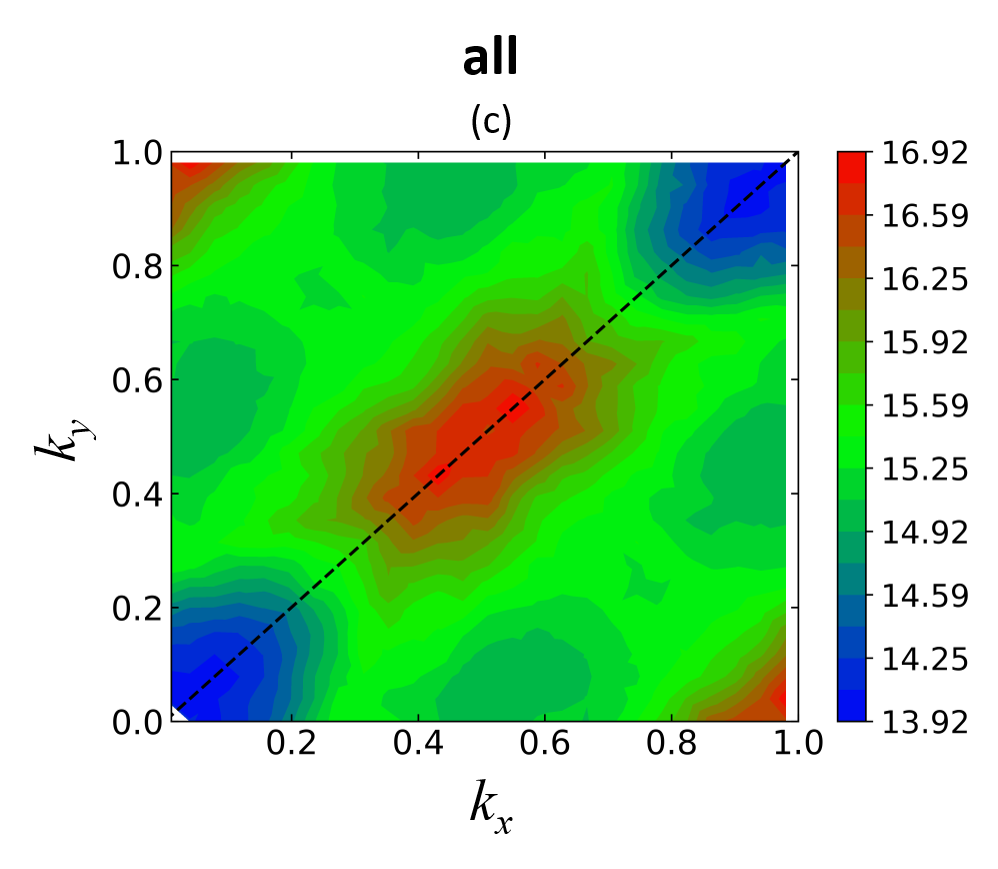}
    
    \includegraphics[scale=0.33]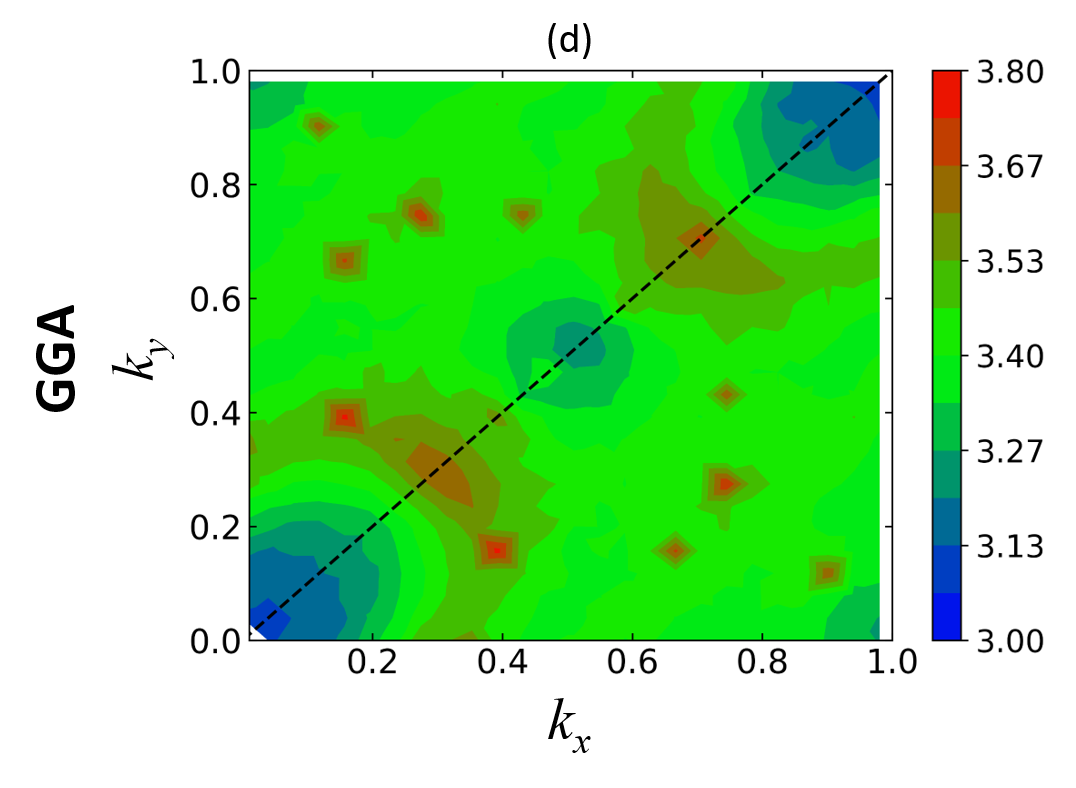}
    \includegraphics[scale=0.33]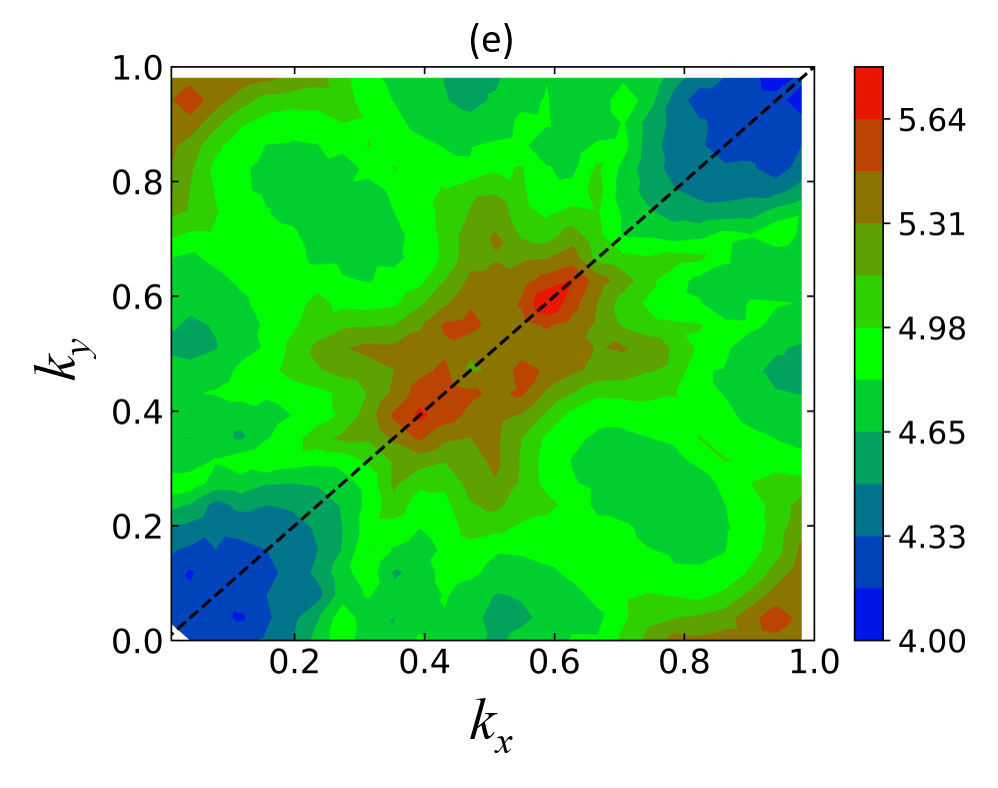}
    \includegraphics[scale=0.33]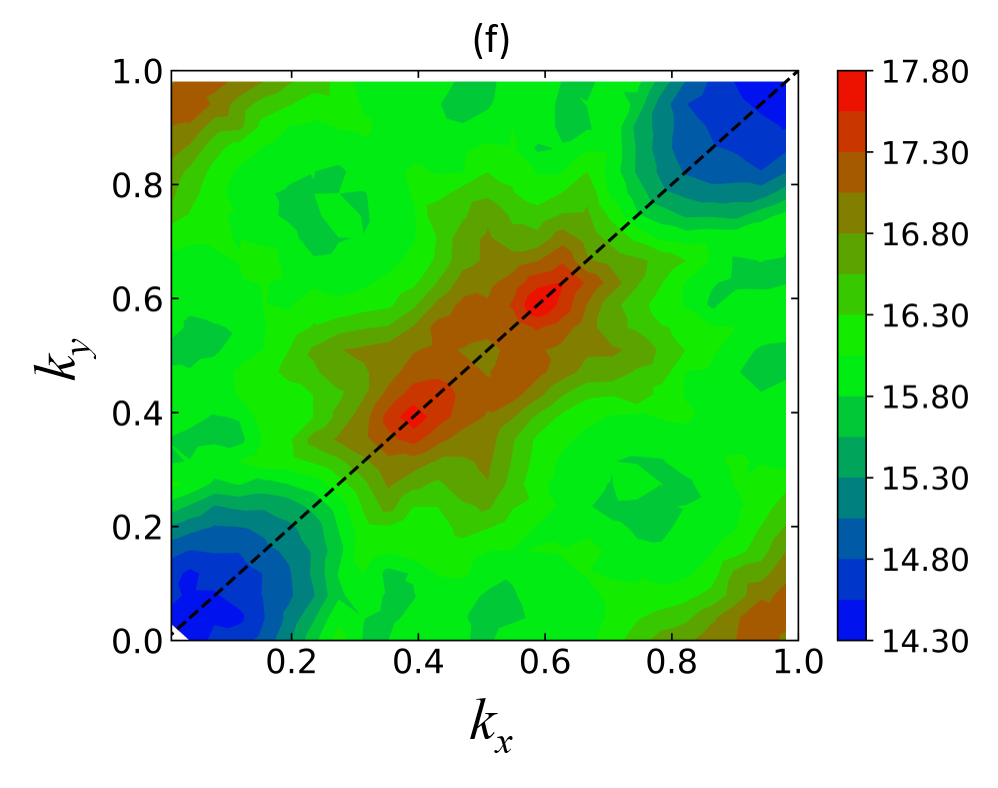}

    \includegraphics[scale=0.33]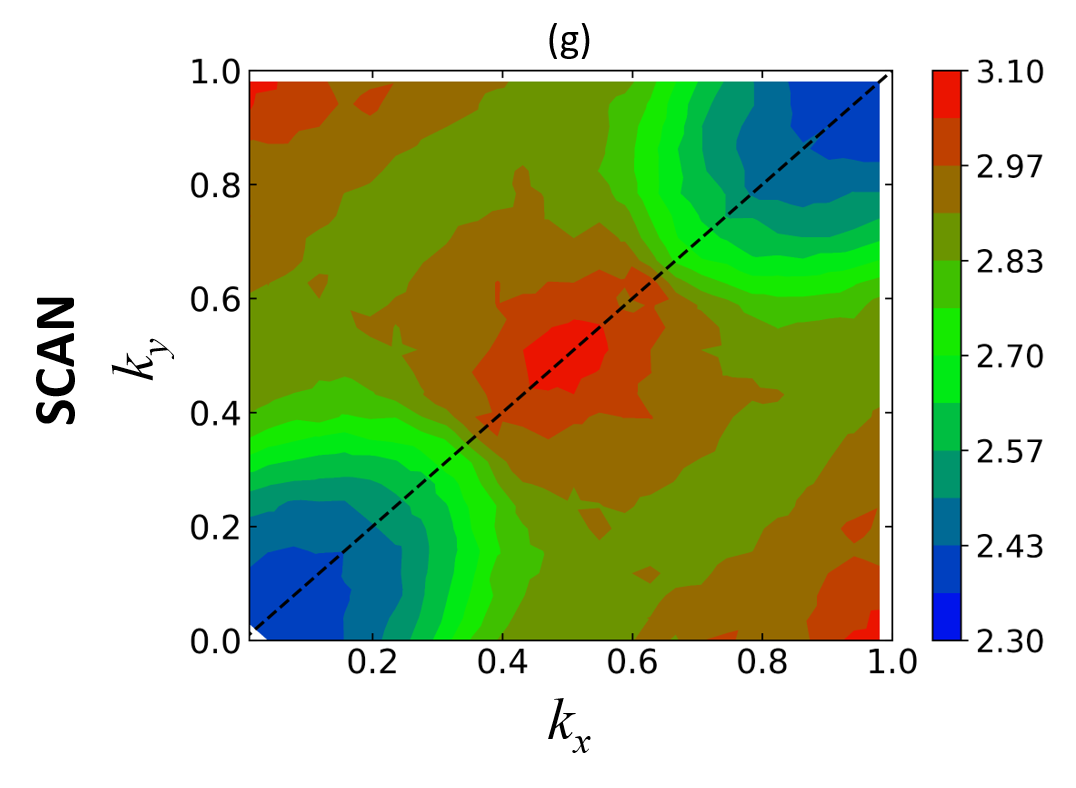}
    \includegraphics[scale=0.33]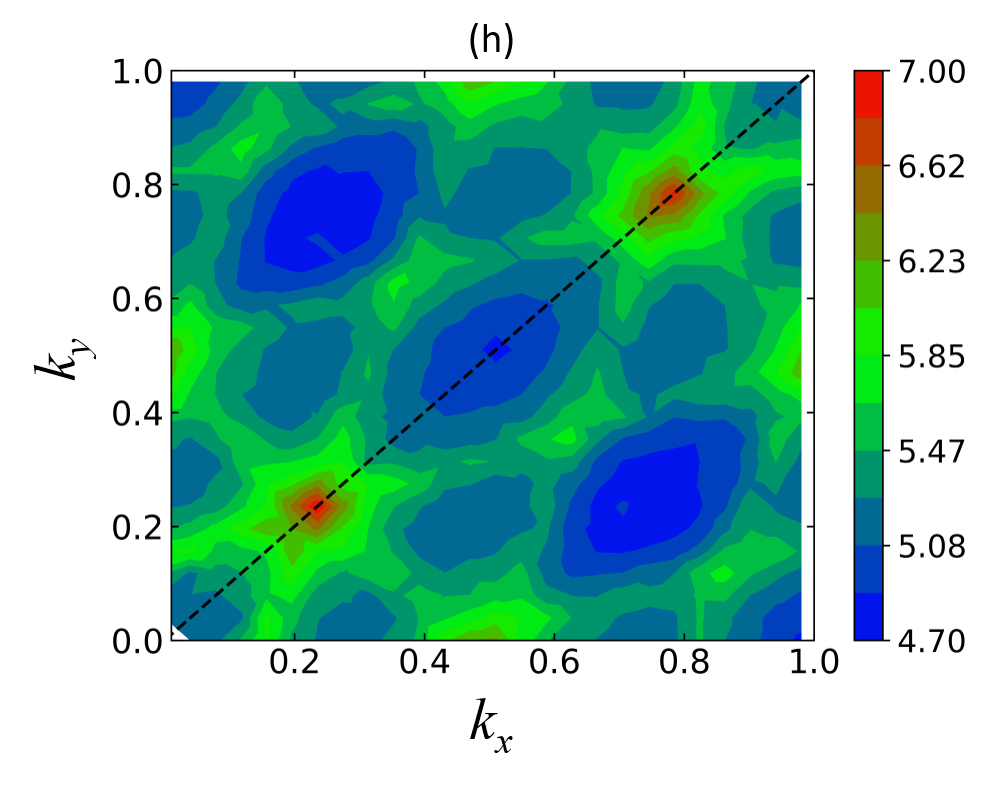}
    \includegraphics[scale=0.33]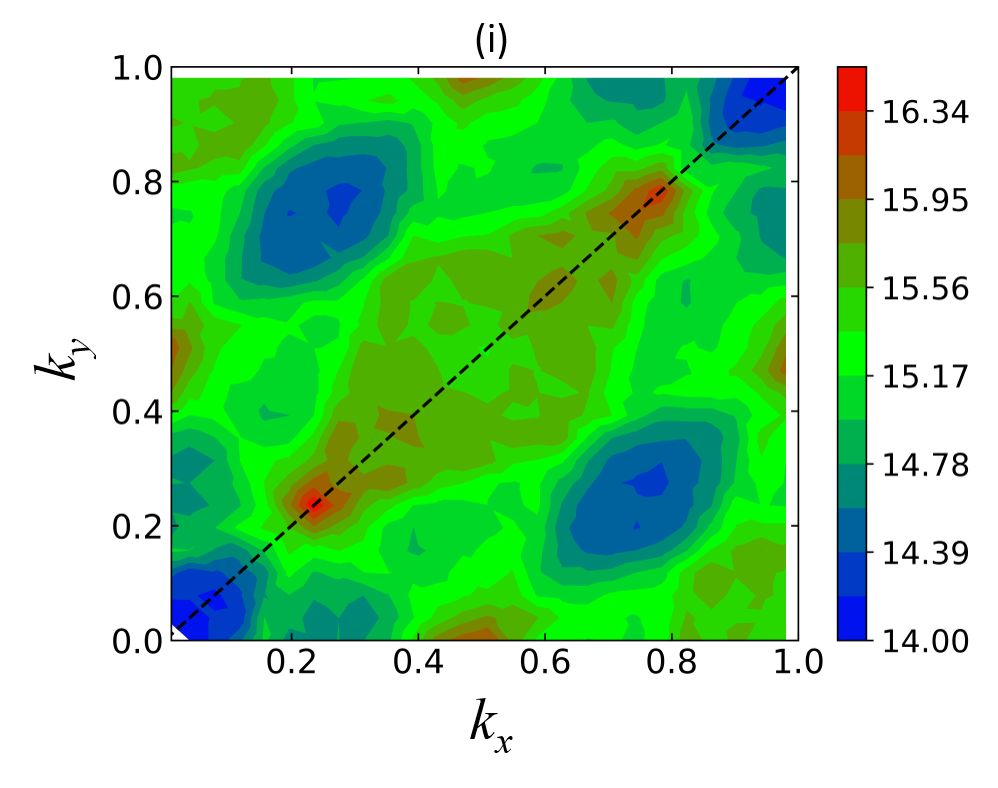}

    \caption{ Generalized susceptibility $\chi$(\textbf{q}) for Ni$_2$MnGa. The top, middle, and bottom panels correspond to LSDA, GGA, and SCAN, respectively. The first, second, and third columns correspond to minority, majority, and total band contributions, respectively.  }
    \label{fig:fig2_s}
\end{figure*}

Figure~S\ref{fig:fig2_s} shows the  2D-maps of $\chi(\textbf{q})$ for the (110) plane calculated with LSDA, GGA, and SCAN. 
The curves for LSDA and GGA show two peaks, which are mainly due to the nesting of the spin-down channel. 
According to the 2D-maps, the results look similar for the minority contributions calculated with LSDA and GGA, while the majority contribution demonstrates some additional peaks of $\chi(\textbf{q})$ in case pf GGA.
SCAN considerable alters the 2D-map. 
In contrast to LSDA and GGA, there is no peak in the major channel for SCAN.
Also, for the total contribution, two additional peaks are observed along the [110] direction.

\section{SCAN$-U$}
\begin{figure}[htb!]
    \includegraphics[scale=0.3]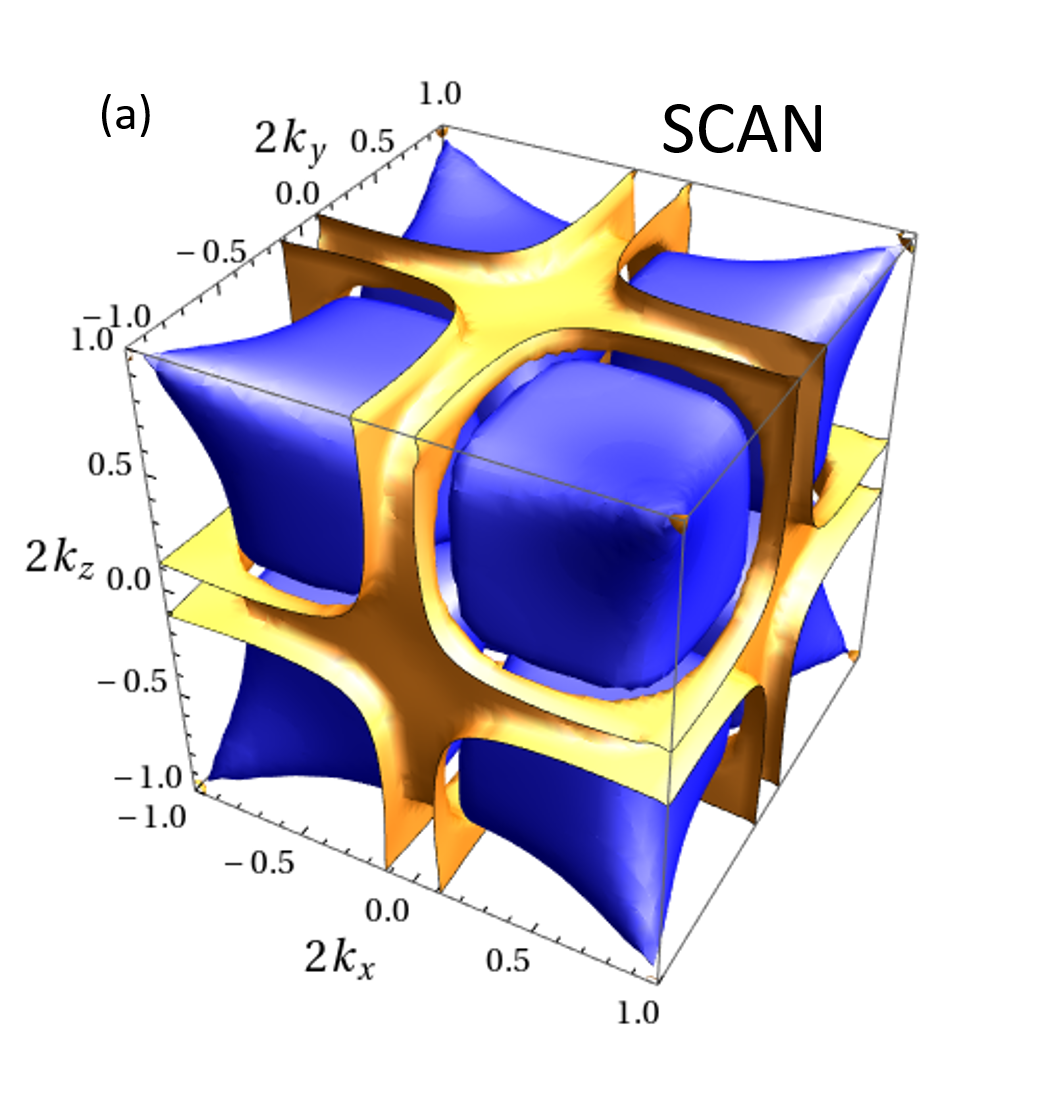}
    \includegraphics[scale=0.3]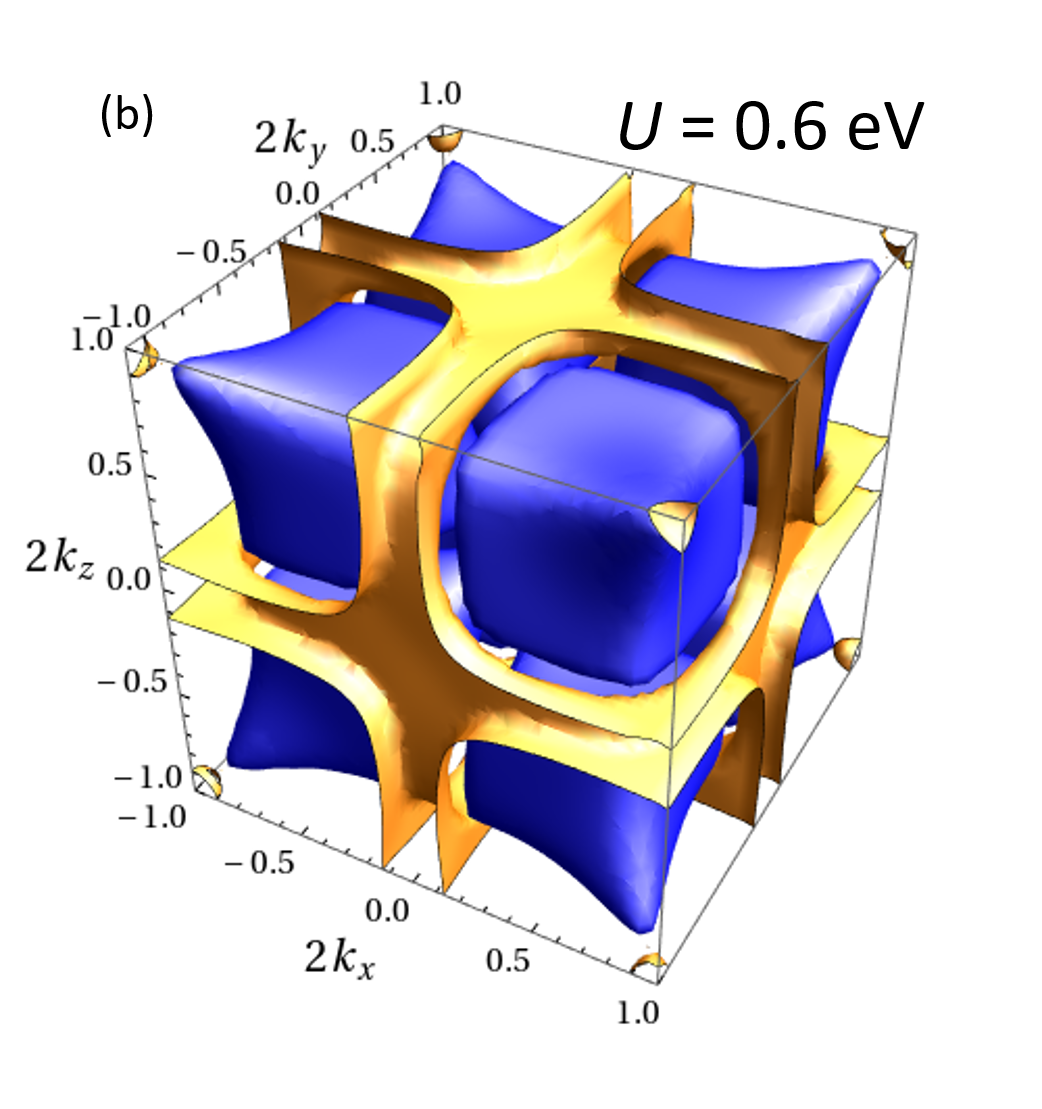}
    \includegraphics[scale=0.3]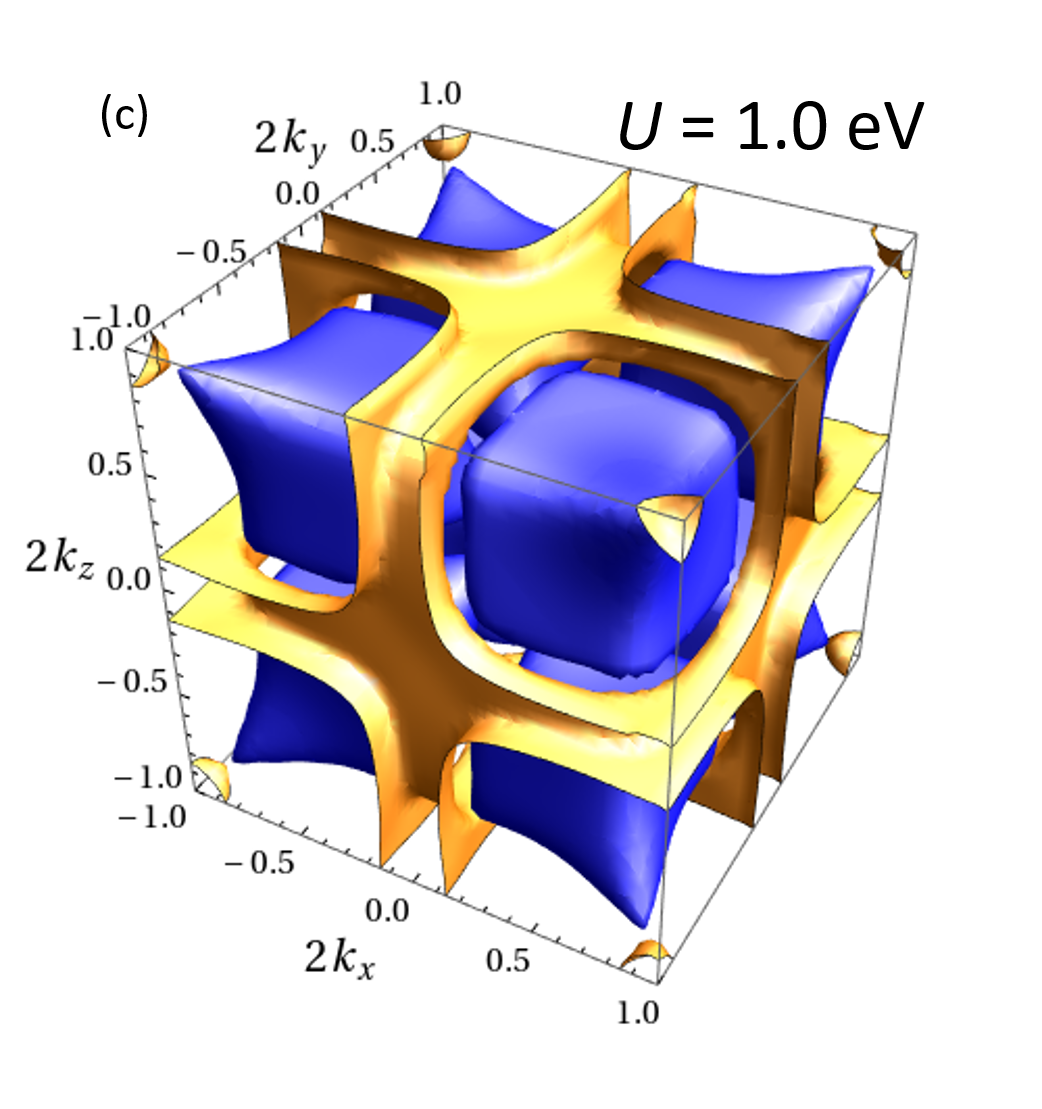}
    \includegraphics[scale=0.3]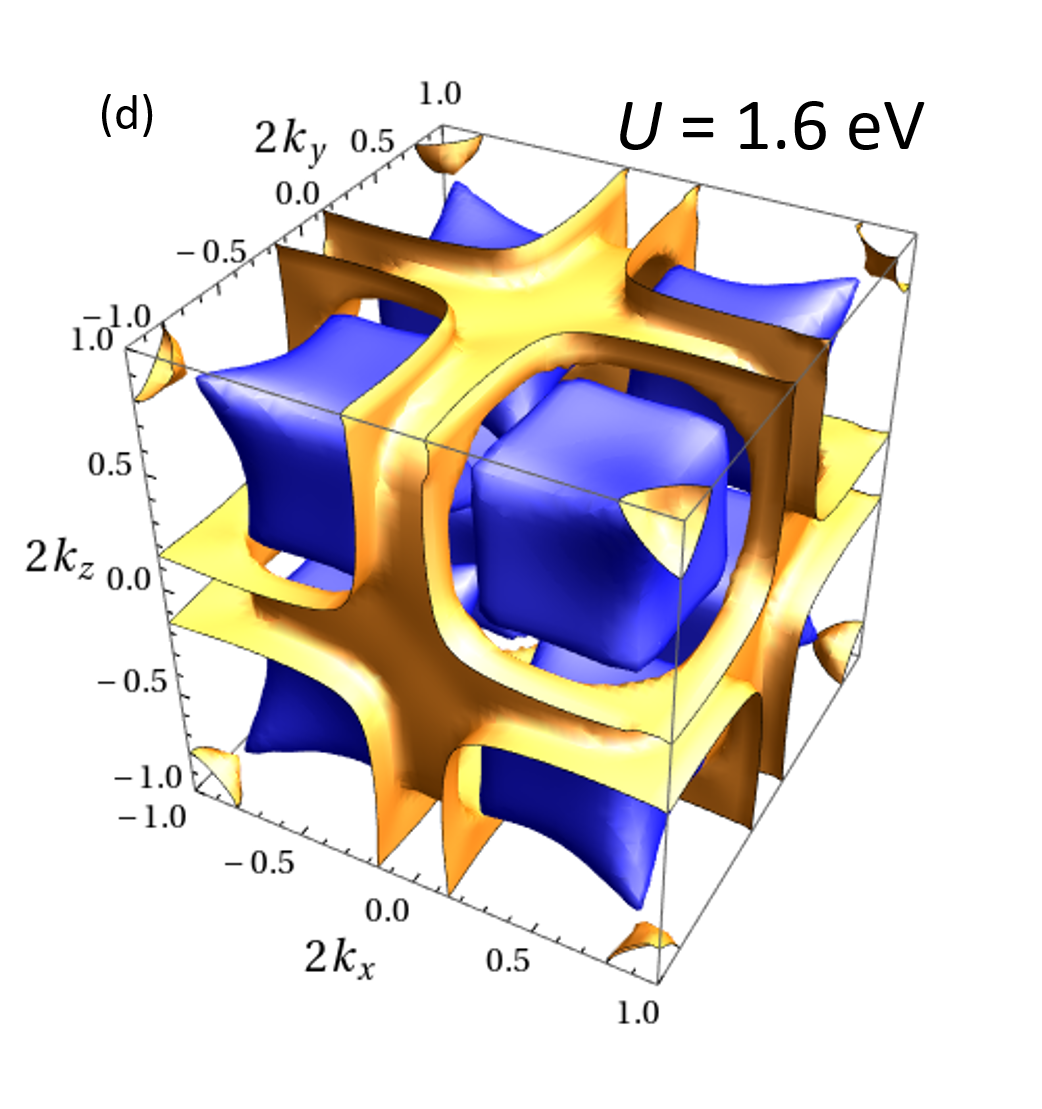}
    \includegraphics[scale=0.3]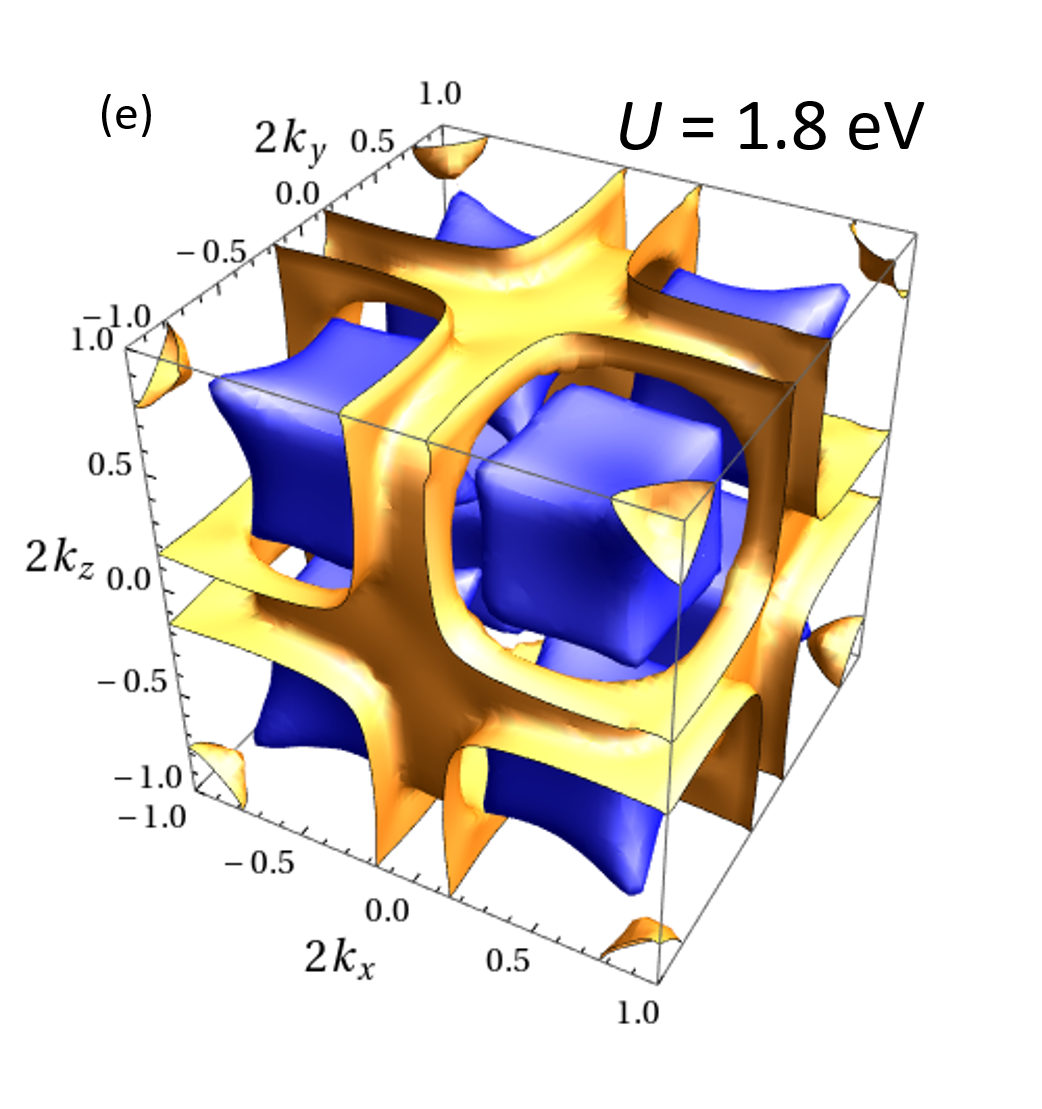}
    \includegraphics[scale=0.3]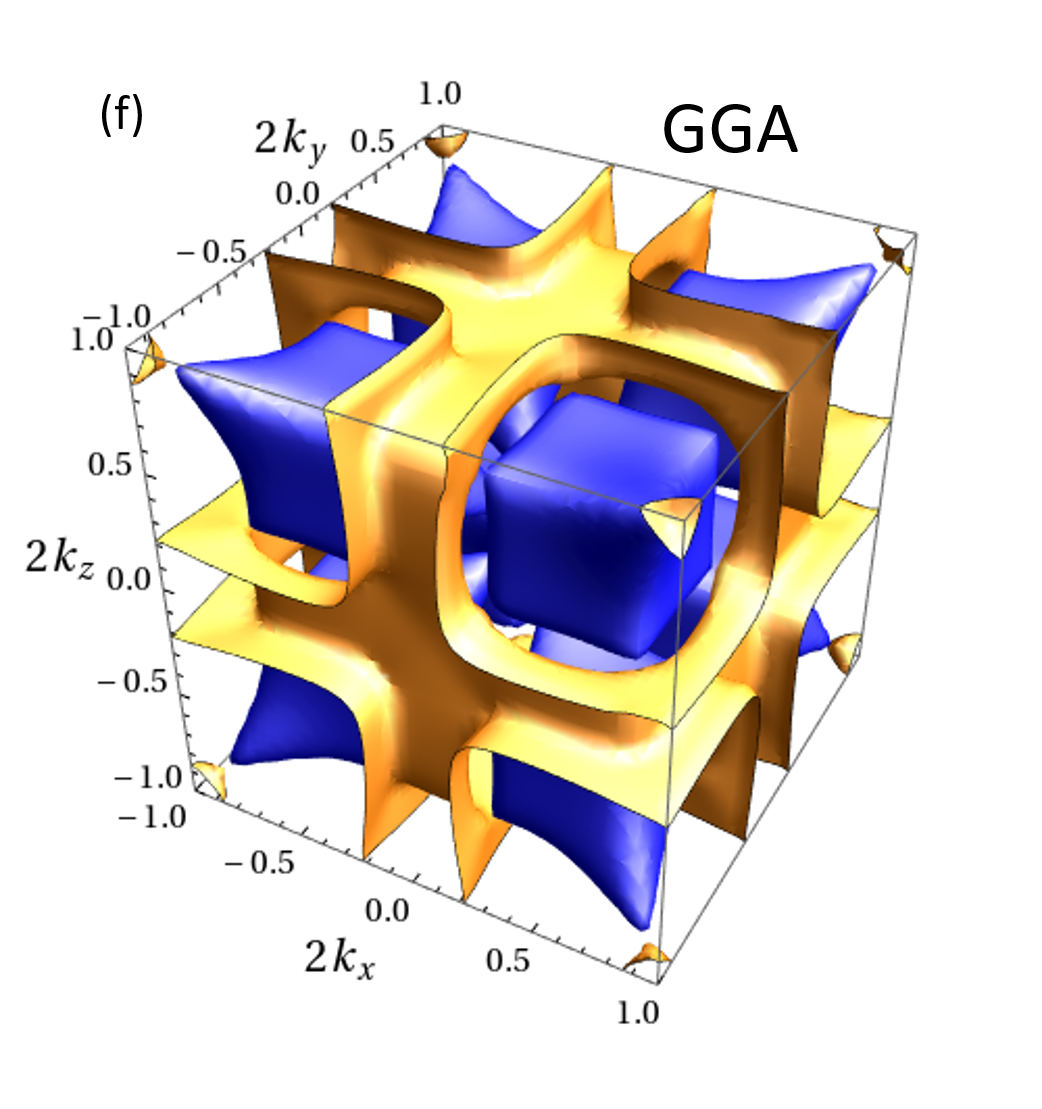}
    \caption{Fermi surfaces calculated with SCAN, SCAN$-U$, and GGA for austenitic phase of Ni$_2$MnGa. }
\label{fig:fig3_s}
\end{figure}

SCAN gives an overestimated magnetic moment \cite{Buchelnikov-2019, Ekholm-2018, Trickey}. 
To reduce it, we use SCAN$-U$. 
The influence of the parameter $U$ on the Fermi surface is illustrated in Fig.~S\ref{fig:fig3_s}.
When $U$ increases, the piece of band~64  indicated by orange expands, while the piece of band~63 indicated by blue becomes narrower. 
This trend leads to an increase in the length of the nesting vector.

\begin{figure}[htb!]
\includegraphics[scale=0.5]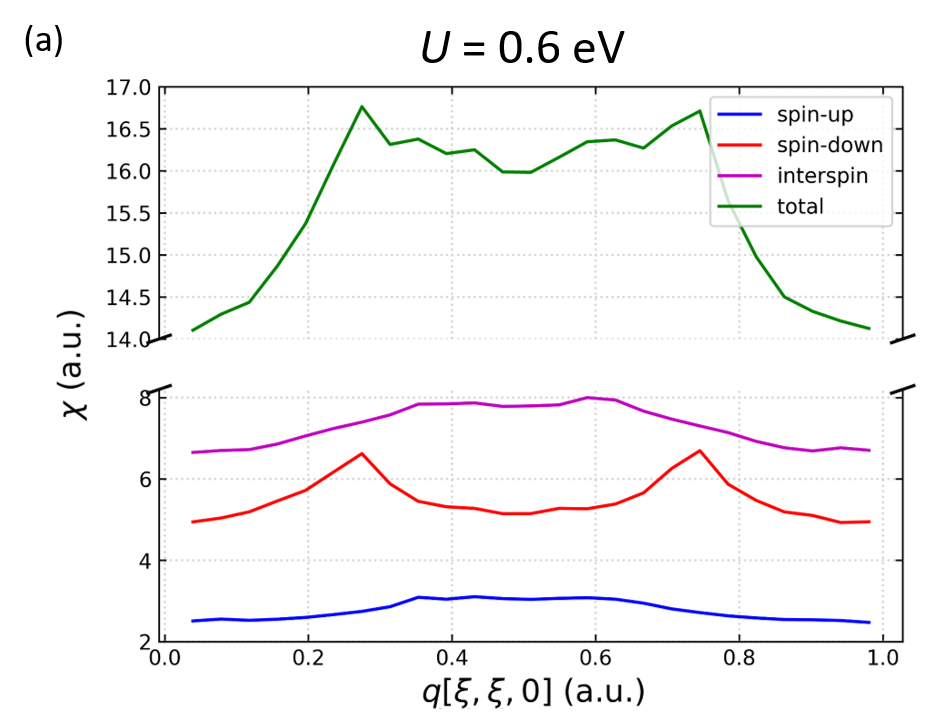}
\includegraphics[scale=0.5]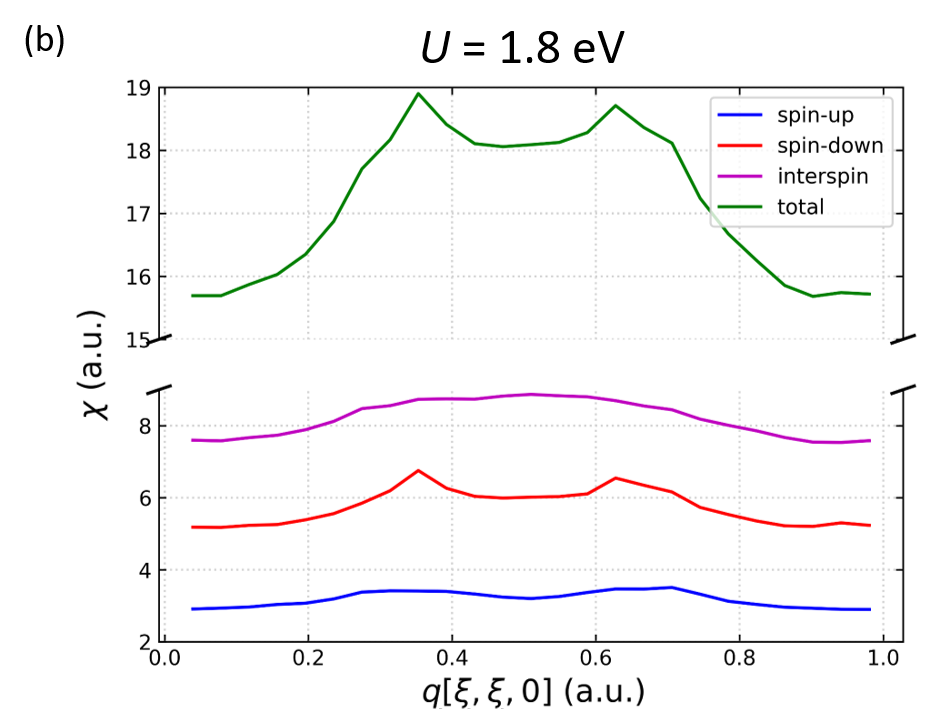}
\caption{ Generalized susceptibility along the [110] direction calculated with SCAN$-U$ for $U=0.6$ and 1.8~eV. The results are presented for spin-up, spin-down, interspin, and total contributions. { \red Spin-up and spin-down are interspin contributions. The total contribution involves both intraspin and interspin contributions in the susceptibility calculation.}}
\label{fig:fig4_s}
\end{figure}

To determine precisely the nesting vector, the generalized susceptibility curves are calculated with different approximations.
The results for SCAN$-U$ ($U = 0.6$ and $U = 1.8$~eV) are presented in Fig.~S\ref{fig:fig4_s}.
In case of $U = 0.6$~eV, the $\chi(\textbf{q})$ for spin-up bands does not show peaks and nesting.
In contrast, $\chi(\textbf{q})$ of spin-down bands shows two pronounced maxima, and the distance between them is slightly lower than in case of SCAN (see Fig.~S\ref{fig:fig2_s}(c)).
For total contribution, $\chi(\textbf{q})$ displays four maxima for SCAN.
However, for $U = 1.8$~eV, the profile of $\chi(\textbf{q})$ becomes similar to GGA. 

\newpage
\bibliography{main}